%% file: paper.tex
\pdfoutput=1 
\documentclass{JINST}
\usepackage{xspace}
\usepackage{subfigure}
\title{Handling uncertainties in background shapes: the discrete profiling method}

\author{P.~D.~Dauncey$^a$\thanks{Corresponding author.},
M.~Kenzie$^b$, N.~Wardle$^b$ and G.~J.~Davies$^a$\\
\llap{$^a$}Department of Physics, Imperial College London, Prince Consort Road, London, SW7 2AZ, UK.\\
\llap{$^b$}CERN, CH-1211 Geneva 23, Switzerland.\\
E-mail: \email{P.Dauncey@imperial.ac.uk}}

\abstract{
A common problem in data analysis is that the functional form, as well as the parameter values,
of the underlying model which should describe a dataset is not known {\it a priori}. In these cases some
extra uncertainty must be assigned to the extracted parameters of interest due to lack of exact knowledge of the functional form of the model.
A method for assigning an appropriate error is presented. The method is based on
considering the choice of functional form as a discrete nuisance parameter which is
profiled in an analogous way to continuous nuisance parameters. The bias and coverage of this method are shown to be good when applied to
a realistic example.
}


\begin{document}
\newcommand{\nll}{\ensuremath{\Lambda}\xspace}

\input introduction.tex
\input concept.tex
\input functions.tex
\input correction.tex
\input conclusions.tex

\acknowledgments
We thank Chris Seez and Louis Lyons for informative discussions.
This work was partially supported by the Science and Technology Facilities
Council, UK.

\bibliographystyle{JHEP}
\bibliography{paper}

\end{document}

%% file: introduction.tex
\section{Introduction} 
\label{sec:introduction}


A common problem in data analysis is that the underlying physics parameters of a model, or components of it, which is used to describe a dataset
are not known. In high energy physics, determination of signal parameters is
often achieved using a maximum likelihood technique~\cite{ref:Fisher01011922}
in which parametric models for the signal and background processes are constructed and
``fit'' to the data. However in certain circumstances the exact parametrisation, not just the parameter
values, of the underlying models is not {\it a priori} known. Consequently, there is some uncertainty in the signal parameters which results from the uncertainty in the function used.

A common approach to assess this systematic uncertainty is to fit various different plausible functions and
determine the spread of the values of the  parameters of interest when using these functions.
However, these methods tend to have some degree of arbitrariness and so
a new approach is discussed in this paper.
This new method was developed as part of the analysis of data at the CMS experiment
following the discovery of the Higgs
boson~\cite{ref:introduction:atlasdis,ref:introduction:cmsdis}.
It was applied to the analysis of Higgs decays to two photons, which
results in a narrow signal on a large
background~\cite{ref:introduction:legacy}.

The method presented is less
arbitrary and treats the uncertainty associated with the
background parameterisation in a way
which is more comparable with the treatment of other
uncertainties associated with the measurement; the choice of background
function results in a systematic uncertainty
which is handled as a nuisance parameter~\cite{ref:intro:nusiances}.
There are two major new components to this approach, namely the method for
treating the choice of function as a nuisance parameter, and how to compare
functions with different numbers of parameters.

The concept is described in Section~\ref{sec:concept}.
The application of the method to functions with the same number of parameters
is described in Section~\ref{sec:functions} and to functions with different
numbers of parameters in Section~\ref{sec:correction}. Further discussion on
the method, namely its practical application to the real-world problem of
the Higgs boson measurements, is given in Section~\ref{sec:conclusions},
along with the conclusions.

Within this paper, twice the negative of the logarithm of the likelihood ratio 
function is denoted by \nll. The data are binned and the
likelihood model used for each bin is the Poisson likelihood ratio to the best
possible likelihood given the observed data, i.e.
\begin{equation}
\nll = 2
\sum_{i} \nu_i - n_i + n_i \ln\left(\frac{n_i}{\nu_i}\right),
\label{eqn:introduction:def2NLL}
\end{equation}
where for the $i^{\rm th}$ bin,
$n_{i}$ is the observed and $\nu_{i}$ is the expected number of events
given a particular background model.
For the purposes of fitting and generating datasets, the statistical package
``RooFit'' is used throughout this paper~\cite{ref:roofit}.


%% file: concept.tex
\section{Concept of the method} 
\label{sec:concept}

When studying a small signal sitting on
a large background, relatively small changes to the background shape can
have significant effects on the apparent signal size and position.
Unless there is a
reliable theory or simulation which can constrain the background shape,
there is an uncertainty about which function to use to parameterize 
the background. Hence, different choices of the functional form
will give different results for the signal parameters, meaning there is a
systematic uncertainty associated with the choice of function.
The basic concept discussed in this paper is to consider this choice
as a source of systematic error which is modelled as a nuisance parameter.
It is then consistently treated like other nuisance parameters as far as
possible.

The two main parameters of a resonant signal are the position (i.e.~the mean)
and size (i.e.~the amplitude), although other
quantities such as the width may also be of interest. This results in a
multi-dimensional parameter space. For simplicity in this paper, the position
and width of the signal are considered to be known and only the signal
size is to be determined. However, the method is also applicable to
cases where more than one parameter is being estimated.

Although the case presented here is searching for a narrow signal 
on a large background, the method is in principle applicable to any data 
analysis which contains unknown components of a discrete nature.

\subsection{Continuous nuisance parameters}
\label{sec:concept:continuous}

It is useful to consider briefly the usual way in which nuisance
parameters are used to incorporate systematic errors. Consider measuring a signal
property for a known background functional shape, but with unknown function
parameters. The background parameters themselves are considered as
nuisance parameters, since their actual values are not of interest.
When fitting for a parameter of interest $x$, \nll
is minimised with respect to $x$ and all of the
nuisance parameters. It is usual to construct a profile likelihood
such that \nll is minimised with respect to the nuisance parameters
for each value of $x$ over a range.
The (for example) 68.3\% confidence interval of $x$
is then
taken as the region for which \nll is less than ${\rm \nll}_{BF}+1$,
where ${\rm \nll}_{BF}$ is the best-fit value at the overall minimum.
Similarly, the 38.3\%, 95.4\% and 99.7\% confidence intervals are defined by the
regions in which \nll is less than ${\rm \nll}_{BF}+0.25$, ${\rm \nll}_{BF}+4$
and ${\rm \nll}_{BF}+9$, respectively.

\begin{figure}[tbp]
\centering
\includegraphics[width=0.6\textwidth]{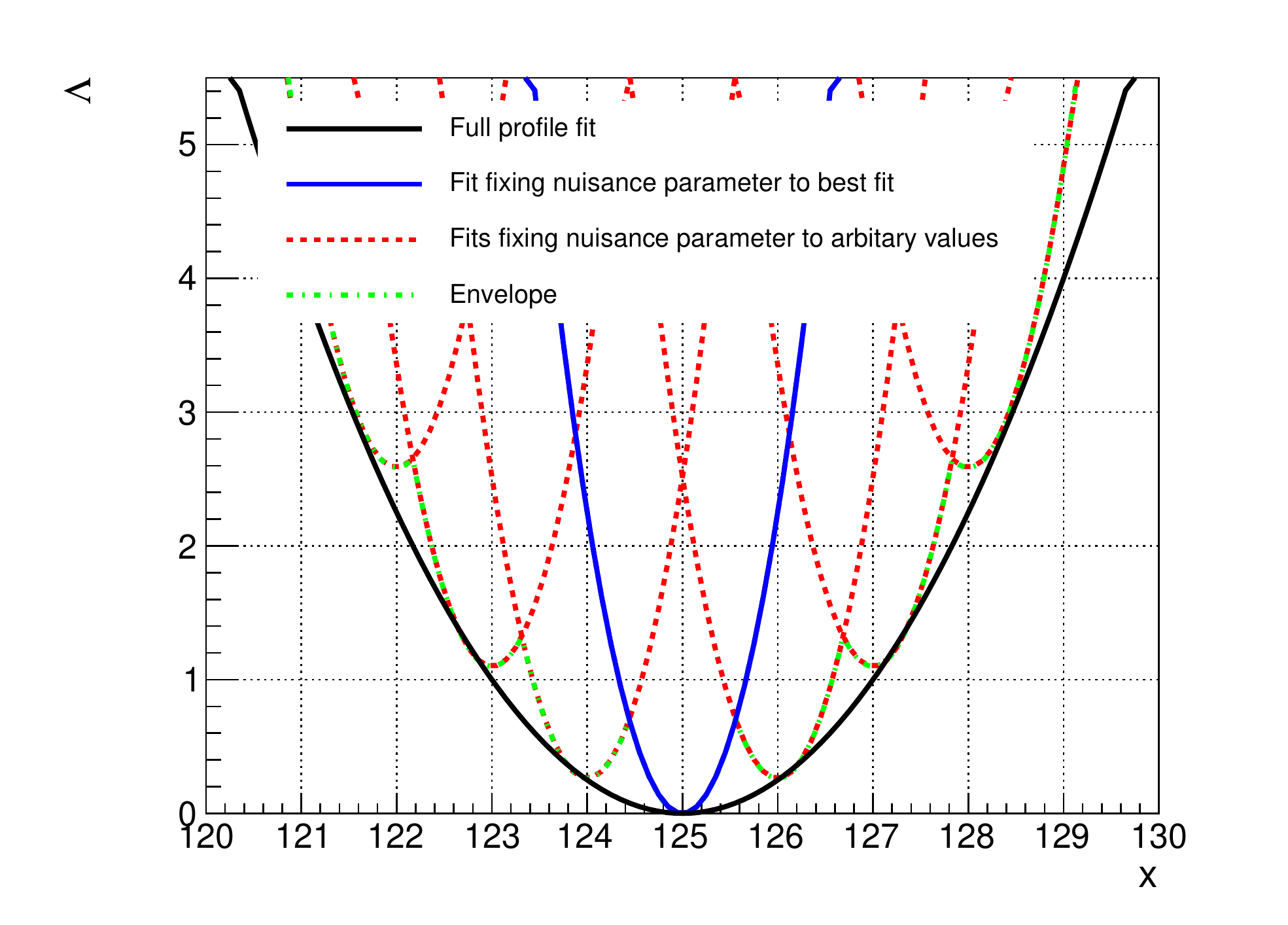}
\caption{Illustration of construction of the envelope (green, dot-dashed line)
by choosing several fixed values of the nuisance parameters (red, dashed lines)
when performing a profile likelihood scan for a variable of interest $x$.
The \nll profile curve for the nuisance parameters fixed to the best fit
values is shown by the blue, solid line. The full profile curve allowing the
nuisance parameters to be fitted for every value of $x$ is shown by the
black, solid line. The red dashed lines show the \nll curves for fixed
nuisance parameter values other than those at the best fit, while the green
dashed line is the envelope, i.e.~the lowest value of any of the red dashed
curves for each $x$. Even with such a coarse sampling of the nuisance
parameters, it is seen that the envelope approximates the full profile curve.}
\label{fig:concept:cartoon}
\end{figure}

An example of a profile likelihood curve for some parameter of interest $x$,
which depends on some nuisance parameters,
is shown by the solid black line in figure~\ref{fig:concept:cartoon}. If the
nuisance parameters were fixed to their values at the best fit point (i.e.~the
absolute minimum in \nll) then the profile \nll curve would be narrower.
This directly reflects the fact that
if the nuisance parameters had no uncertainty,
the error on $x$ would be reduced, i.e.~there would be no systematic
uncertainty arising from this source and the width of this curve would only
be affected by the statistical power of the dataset being fitted.
The same is true for any other values of the nuisance parameters; if their values
were fixed, each would result in a new narrow profile curve with a minimum above the original
curve. The critical point is that the overall minimum as a function
of $x$ encloses the curves for all possible values of the nuisance
parameters.

Now consider finding the profile curve in a different way, illustrated
in figure~\ref{fig:concept:cartoon}.
Each set of nuisance
parameter values will result in a curve, the steepness of which reflects
the statistical error only. Consider picking many different sets of nuisance
parameter values; each gives one such curve. Drawing an ``envelope'' around the
minimum \nll value of all these curves will give an approximation to the
original profile curve. Clearly, in practice, many such sets of nuisance
parameter values would be needed to make this curve smooth. However, in
principle, sampling the nuisance parameter space sufficiently
and then finding the enclosing
minimum envelope should give the profile curve required.
Note, it would of course be possible to mix the two methods, i.e.~choose
sets of only some of the parameter values and fit for the others.

\subsection{Background function uncertainty}
\label{sec:concept:functions}

The method of picking various sets of nuisance parameters and finding
an envelope can conceptually be applied to the uncertainty on the functional form of the background.
Each background function considered can be labelled by an
integer. The integer is then treated as a (discrete) nuisance parameter
for the profile calculation and hence this method has been termed
``discrete profiling''.
The minimum envelope then gives the overall profile curve including
all functions considered. This automatically includes any systematic error
arising from the choice of function, as the envelope will in general be broader
than any of the individual curves which contribute to the envelope.

In practical terms, most minimisation routines
cannot easily handle a discrete parameter, particularly when the
number and values of the other parameters of the fit (i.e.~the parameters
of the functions) change with the discrete parameter.
Hence, in practice it has
to be handled by mixing the methods as discussed in the previous section,
i.e.~as a discrete set of continuous minimisations (one for each function in turn).

While the concept is straightforward, there are some statistical
issues in applying it.
Firstly, there is a question of whether the \nll minimum envelope has the correct coverage,
i.e.~the properties desired for a profile curve.
This is discussed in Section~\ref{sec:functions}.
Secondly, obtaining the envelope profile curve requires a
comparison between the absolute \nll values obtained from fits with different functions, which
in general can include functions with different numbers of parameters.
Hence, a way to meaningfully compare their \nll values must be found and 
again checked for the correct coverage.
This is discussed in Section~\ref{sec:correction}.

%% file: functions.tex
\section{Using functions with equal numbers of parameters} 
\label{sec:functions}

The simplest application of the envelope method is to the case where all
functions used have the same number of parameters.

\subsection{Function definitions}
\label{sec:functions:function}

The first study presented uses four functions, each of which has two parameters.
These functions are chosen as they, and their higher order equivalents,
are feasible representations of the background shape seen in the Higgs to two photon
analysis. The functions are detailed below; in each case $p_0$ and $p_1$ are
the two parameters.
\begin{enumerate}
\item
``Power law''; $f(x) = p_0 x^{p_1}$.
\item
``Exponential''; $f(x) = p_0 e^{p_1x}$.
\item
``Laurent''; $f(x) = p_0/x^4 + p_1/x^5$.
\item
``Polynomial''; $f(x) = p_0 + p_1 x$.
\end{enumerate}

\subsection{Example case}
\label{sec:functions:example}

In the CMS Higgs to two photon analysis, the discrete profiling method is applied to
the actual data taken by the CMS experiment. The variable used for the fit is the invariant mass of the two photons, $m_{\gamma\gamma}$.
The likelihood fit is performed to the $m_{\gamma\gamma}$ spectrum simultaneously
across different event categories\footnote{Categorising events with different signal to background ratios improves the sensitivities
of the analysis but presents additional complications beyond the scope of this study.}.
However, for the purposes of this
paper, the ``data'' used to illustrate the method are generated by a Monte Carlo
method from a smooth background
function which is similar in shape and magnitude to the
real data obtained in one particular category of that analysis. We emphasise that this is not the
actual data sample so that none of the detailed results presented below can be used to deduce any
properties of the Higgs boson itself.
This dataset was generated in 160 bins between 110 and 150\,GeV in
the mass variable, $m_{\gamma\gamma}$.
It included signal events
generated according to a Gaussian distribution with a normalization of 50.8 events, a mean of 125\,GeV and a
width of 1.19\,GeV. These values are representative of the expected Higgs signal from a single category of the
Higgs to two photon analysis.
In the following, the signal strength results are given in
terms of the relative strength $\mu$,
meaning the ratio of the measured number of the signal events relative
to the expected number.

\begin{figure}[tbp]
\centering
\includegraphics[width=0.46\textwidth]{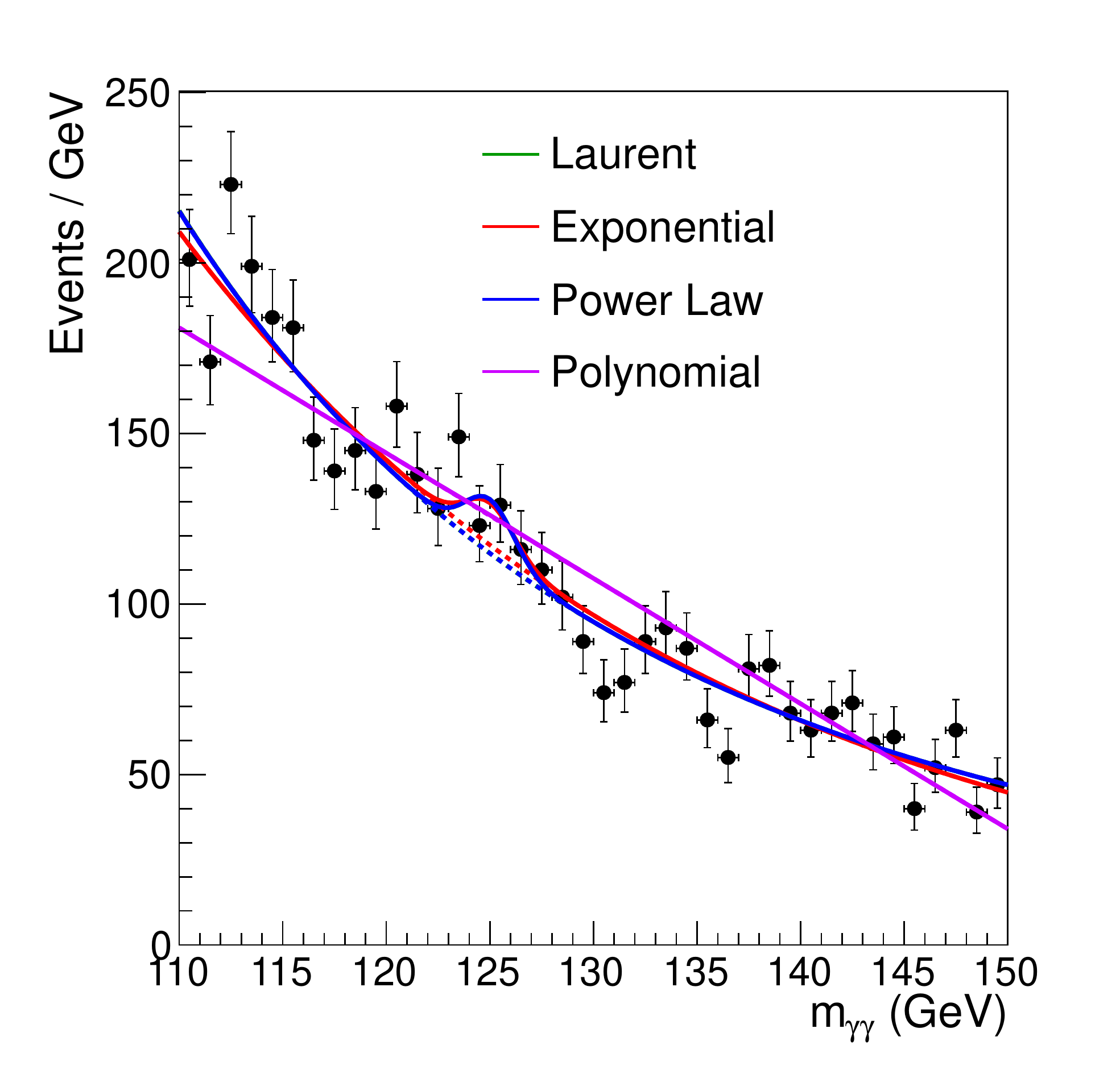}
\caption{Best fits of the four two-parameter functions (described in the
text).
The Laurent function is effectively identical to the power law function
and so is hidden underneath the power law line.
Note, for clarity in this plot, the
data have been rebinned into 40 bins, although the fits were performed with
a finer binning of 160 bins.}
\label{fig:functions:bestfits}
\end{figure}

The four two-parameter functions mentioned above were each
separately fitted to the dataset. A Gaussian signal component was also included, where the
mean and width of the Gaussian were fixed to the same values as used to
generate the events.
The magnitude of the signal Gaussian and both parameters in the
background function were determined from the fit.
The results are shown in figure~\ref{fig:functions:bestfits}.
It is clear that the first order polynomial does not fit particularly well,
while the other three functions give reasonable fits.

\begin{figure}[tbp]
\centering
\includegraphics[width=0.46\textwidth]{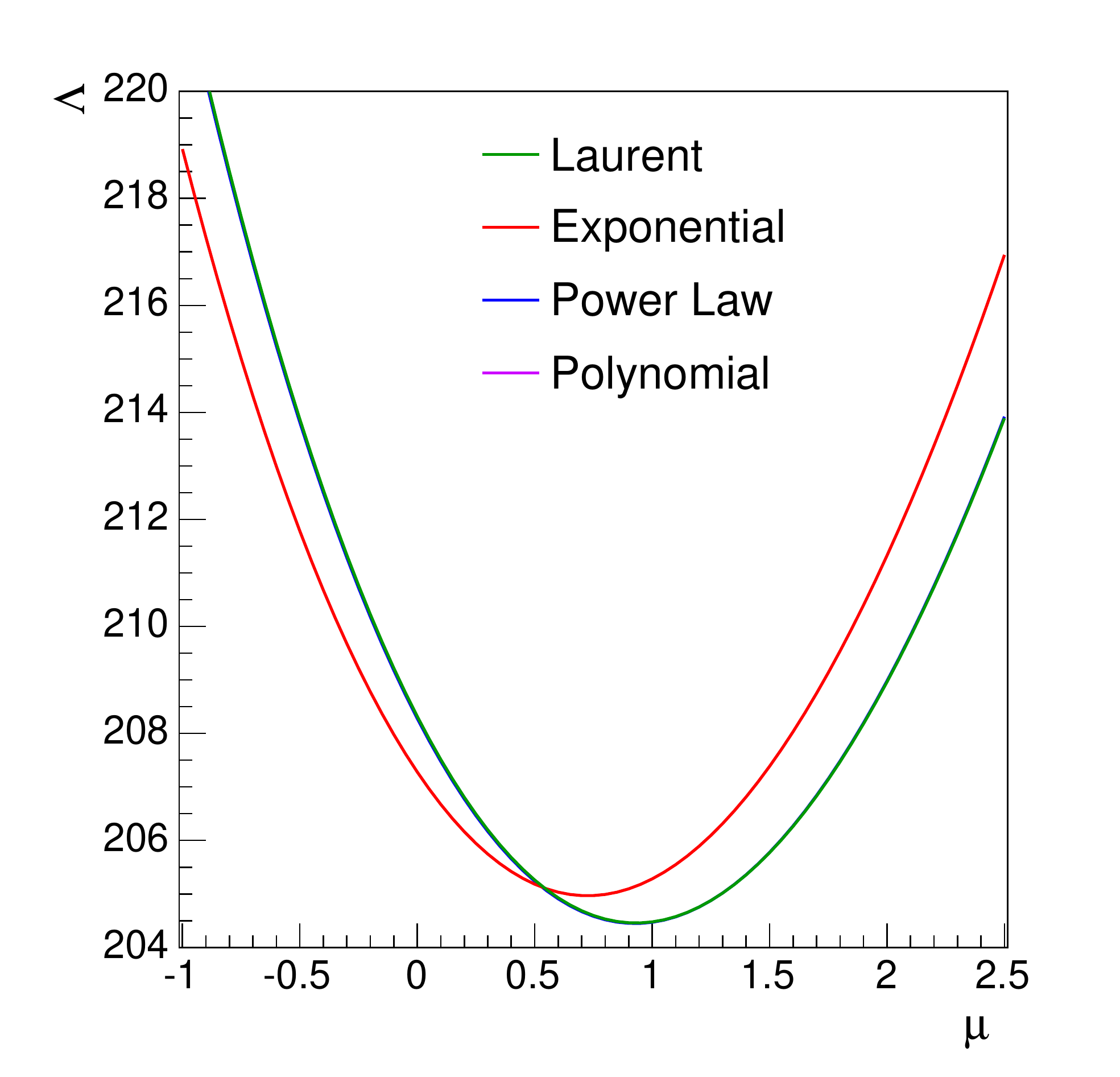}
\caption{Profile \nll scans for the four two-parameter
functions discussed in the text.
The polynomial function is above the top of the \nll scale for all
$\mu$ values shown in this figure. }
\label{fig:functions:profiles}
\end{figure}

The profile scan as a function of the relative signal strength $\mu$
between $-1.0$ and 2.5
for the four functions is shown in
figure~\ref{fig:functions:profiles}.
The absolute minimum occurs for the power law function at a relative signal
strength of $\mu = 0.93$, for which the function parameters have the values
$p_0 = 2.24\times 10^{12}$ and $p_1 = -4.91$.
If only this function is considered,
the 68.3\% confidence interval on $\mu$ is
$0.43 < \mu < 1.40 $, determined as the interval for which $\Delta{\rm \nll} = \nll - \nll_{BF} < 1$.
The measurement would therefore be reported with its standard error as $\mu=0.93^{+0.47}_{-0.50}$. The Laurent function gives an identical result within the precision given.
If only the exponential function is considered, a slightly higher
\nll value is obtained
at the best fit point, which corresponds to a relative signal strength of $\mu = 0.72$, with parameters $p_0 = 1.45 \times 10^4$ and $p_1 = -0.0386$.
The 68.3\% confidence interval is
$0.27 < \mu < 1.24 $, equivalently written as $\mu = 0.72^{+0.52}_{-0.45}$.
Fitting with the straight line yields a very different result of
$\mu = 0.01^{+0.51}_{-0.47}$
although it is clear that this function does not describe the data well and it
gives a much larger \nll value.
The fact that the different functions can give different best fit values
is a direct example of the systematic uncertainty associated
with the choice of function.

\begin{figure}[tbp]
\centering
\includegraphics[width=0.46\textwidth]{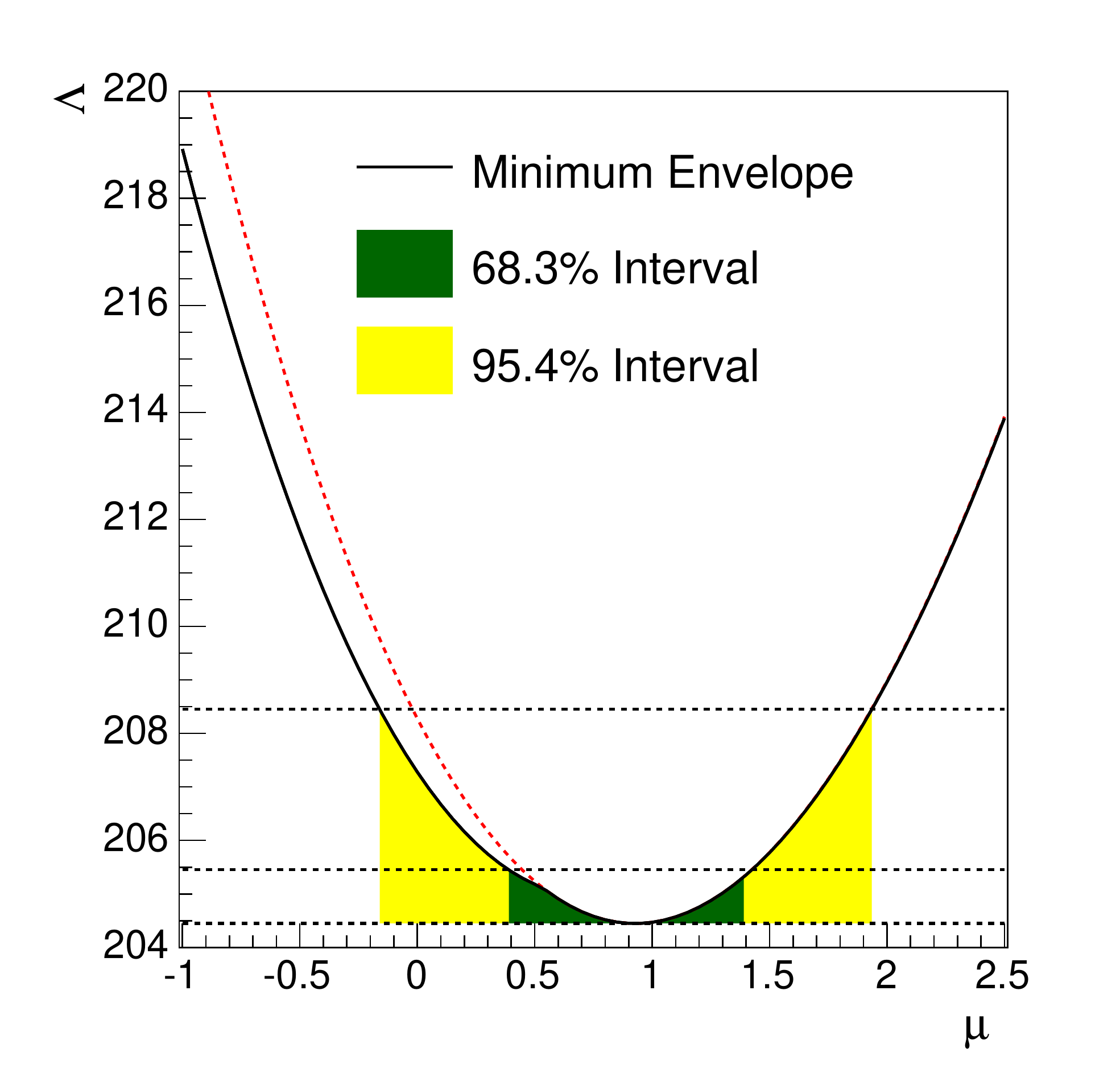}
\caption{Profile \nll envelope for the four two-parameter function fits.
The coloured bands indicate the 68.3\% and 95.4\% intervals determined from the regions
for which the value of \nll increases by 1 and 4 units from the minimum value as indicated by the horizontal lines. The dashed red line shows the profile \nll
curve which would be obtained using just the power law function.}
\label{fig:functions:envelope}
\end{figure}

The envelope around these functions is shown in
figure~\ref{fig:functions:envelope}.
By construction, the best fit is still $\mu=0.93$ from the power law
but now the standard error is enlarged by the contribution from the exponential function
on the lower side of the scan. Hence, taking all four functions into
account, the 68.3\% confidence interval on $\mu$ is
$0.37 < \mu < 1.40 $, i.e.~the lower limit is extended by
the exponential fit
compared with the power law fit alone.
The measured value of $\mu$ would now be quoted as
$\mu = 0.93_{-0.56}^{+0.47}$.
The enlarged uncertainty is a direct reflection of the
systematic error arising from the uncertainty on the choice of function.
As also shown in figure~\ref{fig:functions:envelope}, the 95.4\% confidence
interval is $-0.18 < \mu < 1.92$.
There are two things to note. Firstly, although the Laurent fit
is effectively identical to the power law, there is no issue with
``double-counting'' as the envelope just takes the lowest \nll.
Also, it is clear that the poor fit of the polynomial
means it plays no role in the envelope and so this function is
``automatically'' ignored by the method,
without requiring any arbitrary criterion for
including it or not.

\subsection{Checks of the method}
\label{sec:functions:coverage}

Properties of this method were studied
using a large ensemble of
pseudo-experiments (``toys''). In each pseudo-experiment, a
dataset including signal and background events was generated using a Monte Carlo technique.
Ensembles were generated 
under various different background hypotheses and for different values of the signal strength, $\mu$.
The resulting toy datasets are treated identically to the original dataset.

The background function from which the Monte Carlo events were generated was chosen to be one of the power law,
exponential or Laurent two-parameter functions discussed in
Section~\ref{sec:functions:function}. The background parameters for generating the toys are set to their best fit values for the given value of $\mu$.

In addition, a further ensemble of toy datasets was generated, for which the background function itself, as well
as its parameters, was chosen according to the best fit (i.e.~the function minimising \nll) for each value of $\mu$.
As can be seen from figure~\ref{fig:functions:profiles}, this means that for values of $\mu < 0.55$ the exponential background
function will be used and it will be a power law function otherwise.
Conceptually, this method again
treats the choice of function as a discrete nuisance parameter and so picks the
best fit values of {\em all\/} nuisance parameters for each $\mu$ value.

\begin{figure}[tbp]
\centering
\includegraphics[width=0.46\textwidth]{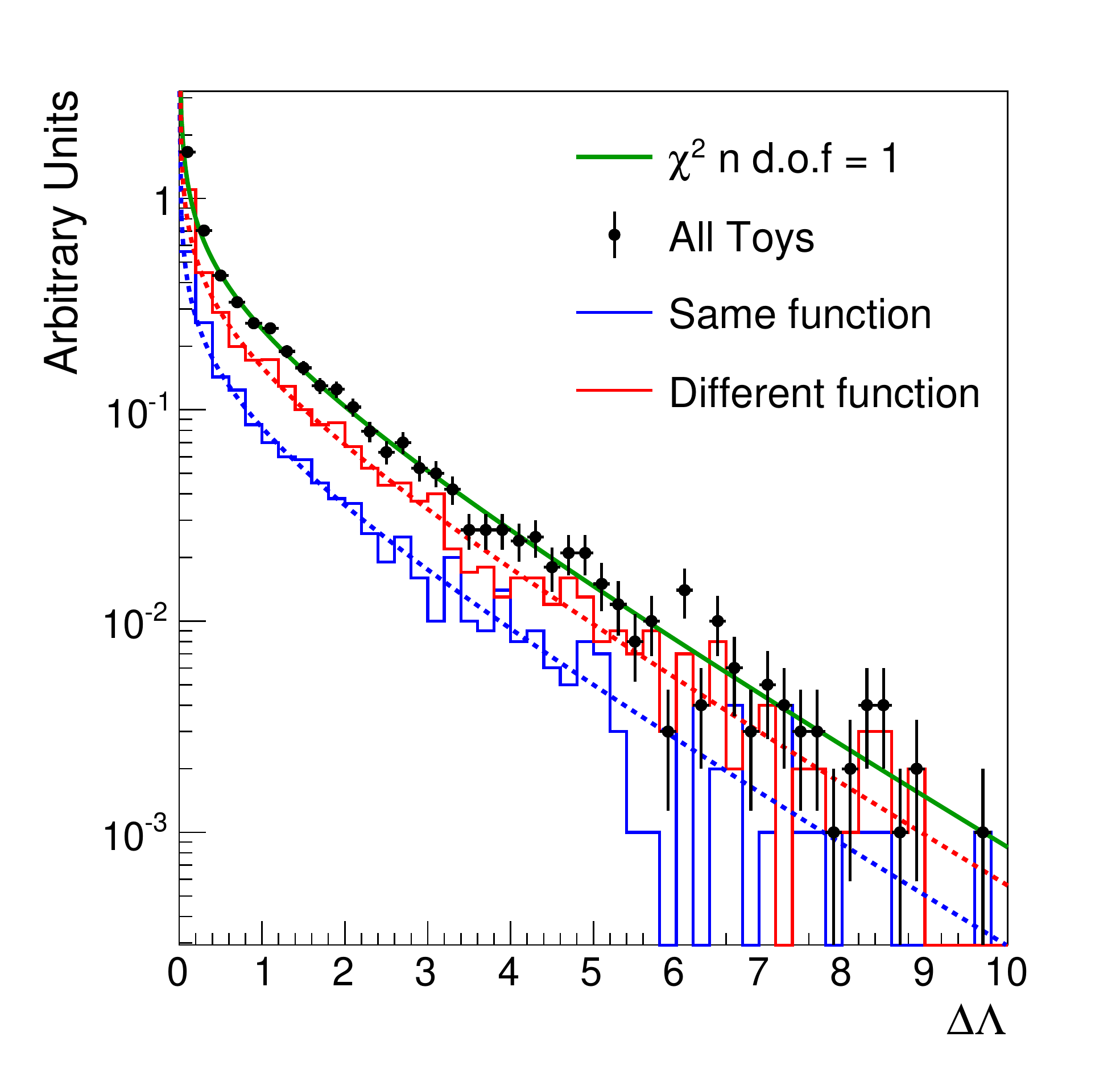}
\caption{Distribution of  $\Delta\nll$, the difference
between the \nll value with $\mu$ fixed to its true value and \nll at the best fit value
of $\mu$. These values are from fits to toy datasets generated with 
$\mu=1$ and with the power law function, for which the parameters are fixed to the best fit values as described in the previous section.
The black data points are from the toy dataset fits and the green function
shows the expected $\chi^2$ distribution for one degree of freedom. The blue and red histograms show the distributions from the toys 
separated into cases where the best fit uses the same or different functions, respectively, compared with the function used to generate the toys. 
The power law
(same function) was the best fit function in 34.1\% of the toys, 
while the Laurent and exponential were the best fit functions in 
36.5\% and 29.4\% of the toys, respectively.
The dashed red and blue lines are $\chi^{2}$ distributions with one degree of freedom normalized to the number of toys in the red and blue histograms respectively.}
\label{fig:functions:chisq}
\end{figure}

The difference $\Delta\nll$  between \nll at the true (generating) value of $\mu$ and at the best fit value is expected to be distributed, in the 
asymptotic limit, as a $\chi^{2}$ with one degree of freedom when considering a single function.
Figure~\ref{fig:functions:chisq} shows an example of the distribution of $\Delta\nll$ in one of the ensemble of toys, for which $\mu=1$ and
the power law is used as the background function
for generating the toy datasets.
The distribution is split into two cases; the toys for which the minimum of \nll is achieved with the 
power law (same function) and the toys for which one of the other three functions provides the minimum \nll.
It is seen that the distribution in toys agrees well with a $\chi^2$ distribution in both cases separately and for the whole ensemble. 
This is true even though the best fit function is more often different from the generated function rather than being the same.
This example is typical; for all the functions used to generate the toy datasets, the distributions of $\Delta\nll$ give the same conclusion.
This indicates that the actual function which contributes to the envelope is not an important factor in obtaining the correct result
and hence treating the function used as a nuisance parameter (i.e.~one for which the value is not important) is a sensible approach. 

\begin{figure}[tbp]
\centering
\includegraphics[width=0.7\textwidth]{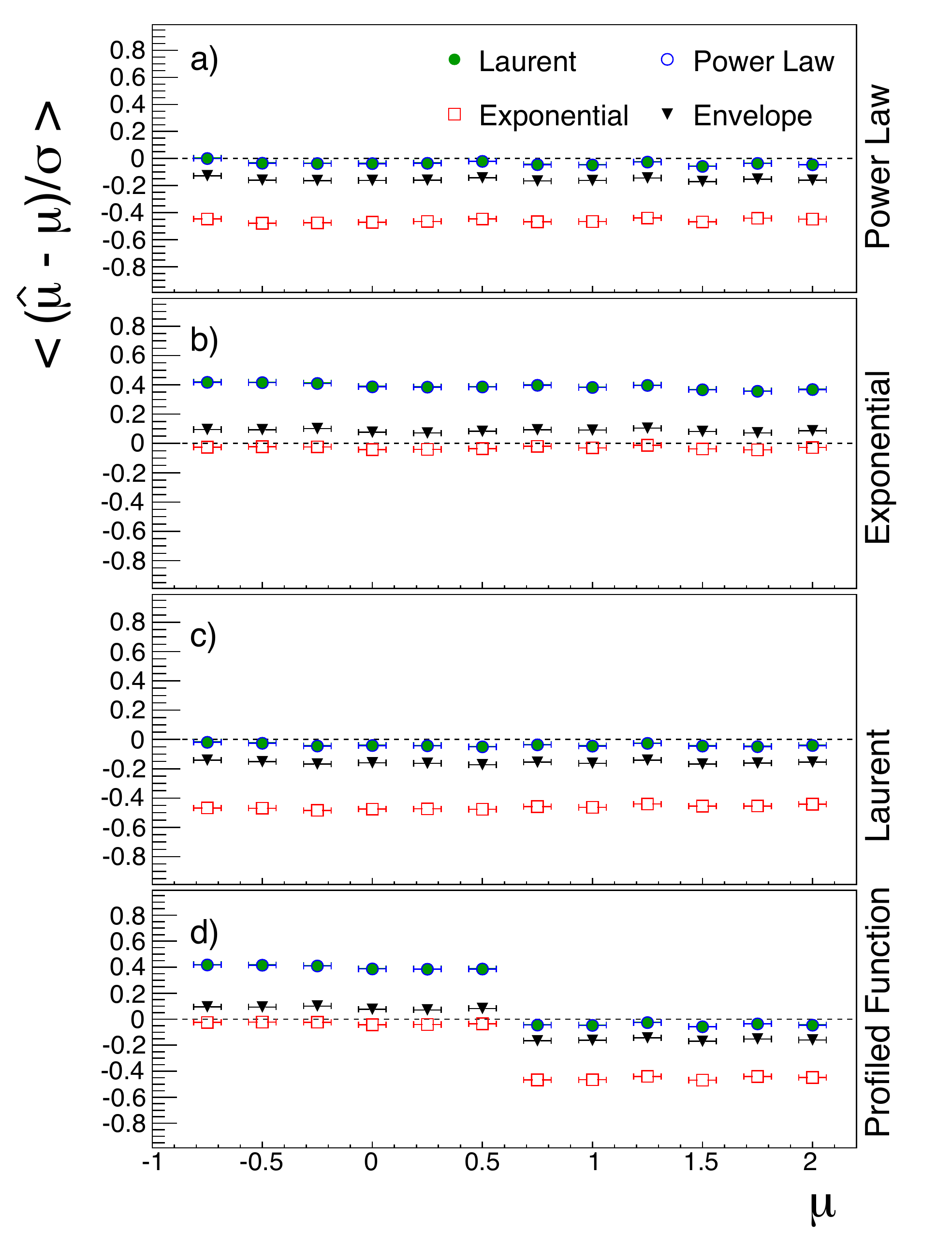}
\caption{Average pull when fitting with each function as background and when
using the discrete profiling method. Panels (a), (b) and (c) show the results
when the generating background function is the power law, exponential and Laurent,
respectively. Panel (d) shows the result when the best-fit function at each
value of $\mu$ is used to generate toys; this means the exponential function
below $\mu = 0.55$ and the power law function above this value. Within each panel the different
points correspond to a different fitting function: Laurent (solid green), power law (open blue), exponential (open red) and the envelope of all four two-parameter functions (solid black). The power law values are effectively identical
to those for the Laurent.
In all cases,
fitting with the polynomial gives values outside the range of these plots.}
\label{fig:functions:firstorderbias}
\end{figure}

The same toy datasets were also used to check the bias and coverage.
These are assessed by calculating the fitted signal strength, $\mu$, and its error, $\sigma$, for each toy when using different background models. The background functions fitted to the toy datasets were the four two-parameter functions discussed.
We define the bias as the mean of the pull distribution, where the pull for an individual toy dataset is defined as
\begin{displaymath}
	p(\mu,\sigma) = \frac{\hat{\mu}-\mu}{\sigma},
\end{displaymath}
where $\mu$ is the generated value of the signal strength, $\hat{\mu}$ is the fitted value of $\mu$ per toy and $\sigma$ is the positive error on $\hat{\mu}$ if $\hat{\mu} < \mu$ and is the negative error on $\hat{\mu}$ if $\hat{\mu} > \mu$. The results of the mean pull as a function of the generated signal strength are shown in figure~\ref{fig:functions:firstorderbias}. It can be seen, as one would expect, that when fitting with the same background function as used to generate the toy dataset, the bias is negligible.
However, when fitting with a different background function the bias can be large; here giving a mean pull up to 0.5.
The discrete profiling method provides a medium between these two in which the bias is small (a mean pull of around 0.1) regardless of the generating function used. This is important given that this method is to be applied when the true underlying function is unknown.

\begin{figure}[tbp]
\centering
  \subfigure[]{\includegraphics[width=0.4\textwidth]{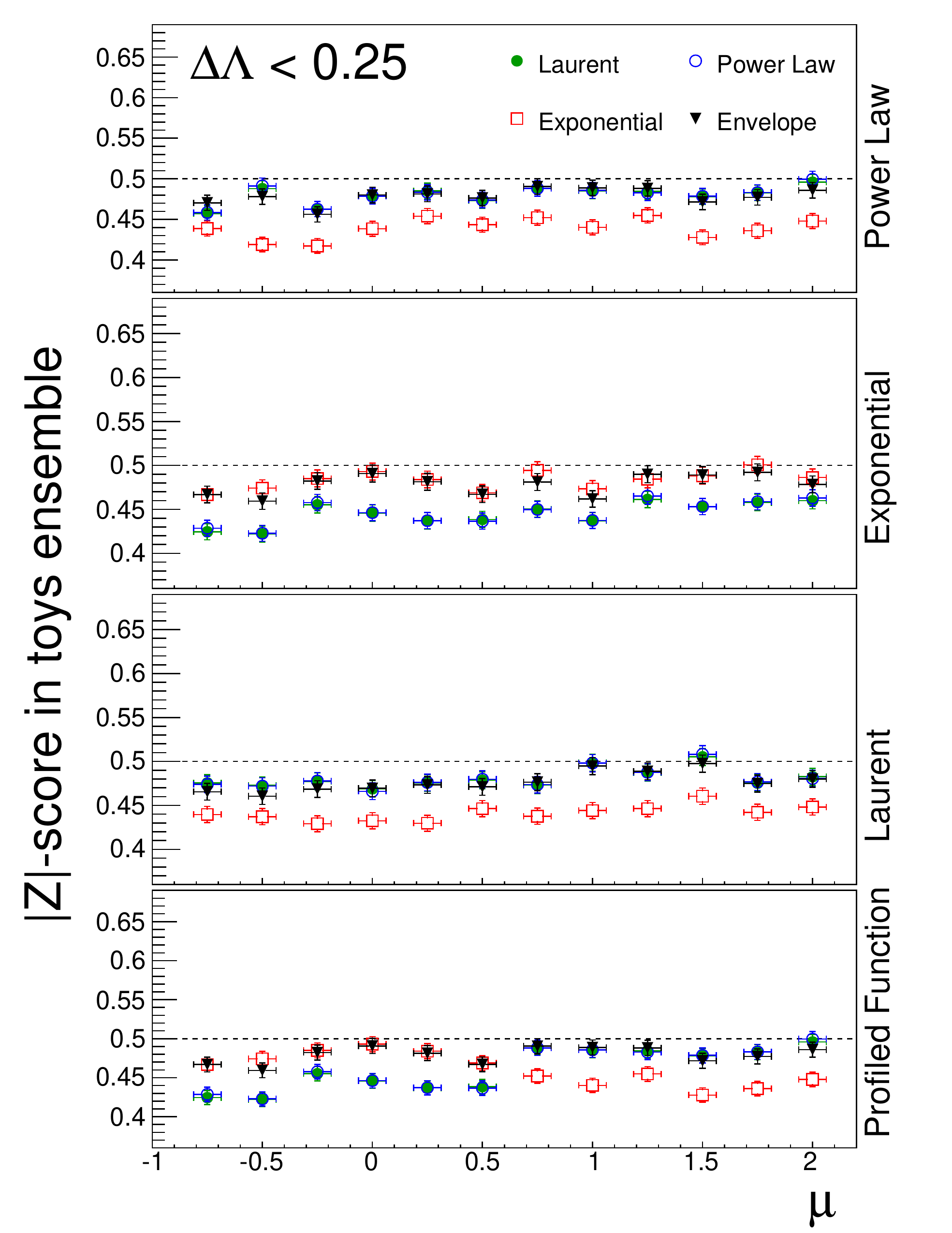}
\label{fig:functions:firstordercoverage:a}}
  \subfigure[]{\includegraphics[width=0.4\textwidth]{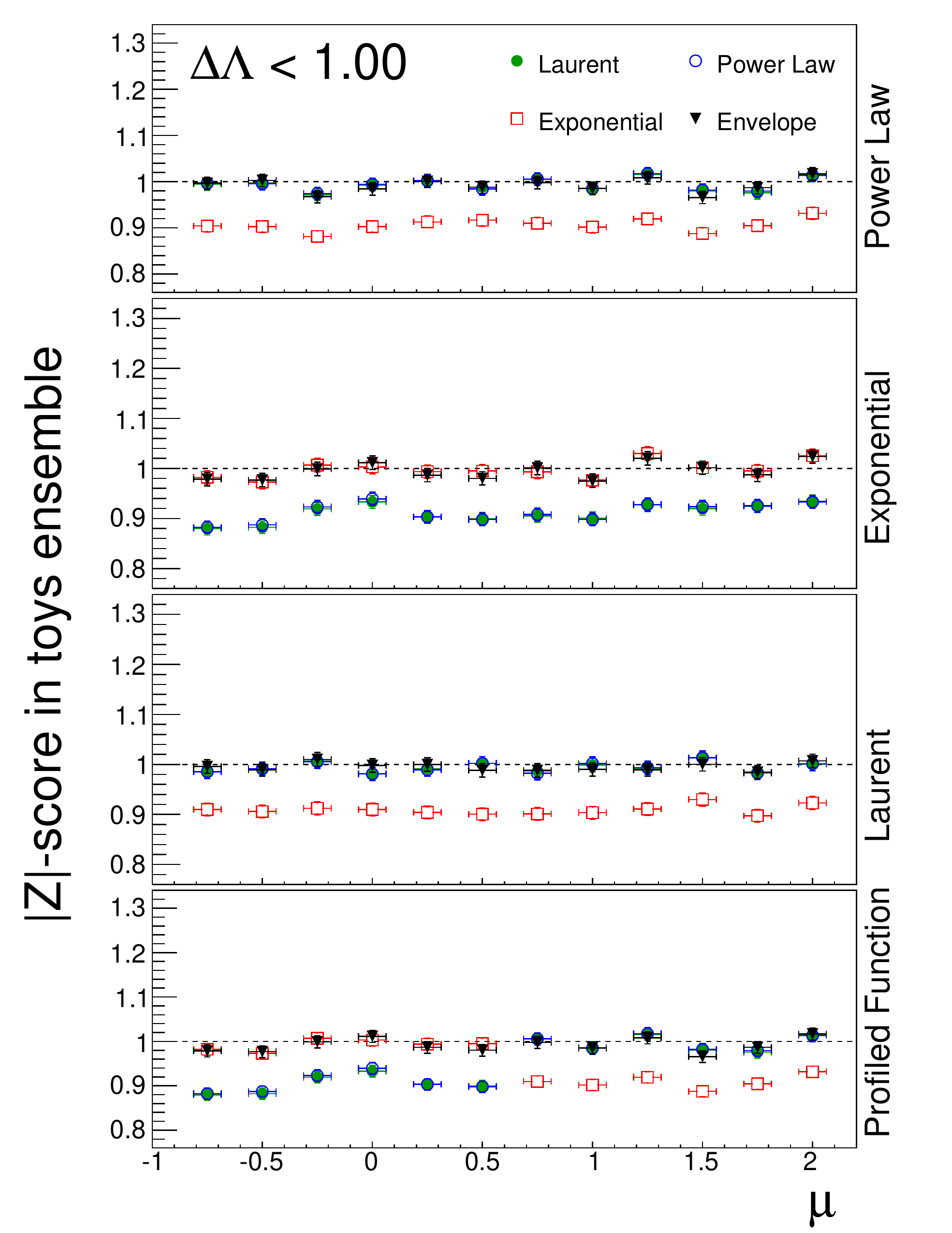}
\label{fig:functions:firstordercoverage:b}}\\
  \subfigure[]{\includegraphics[width=0.4\textwidth]{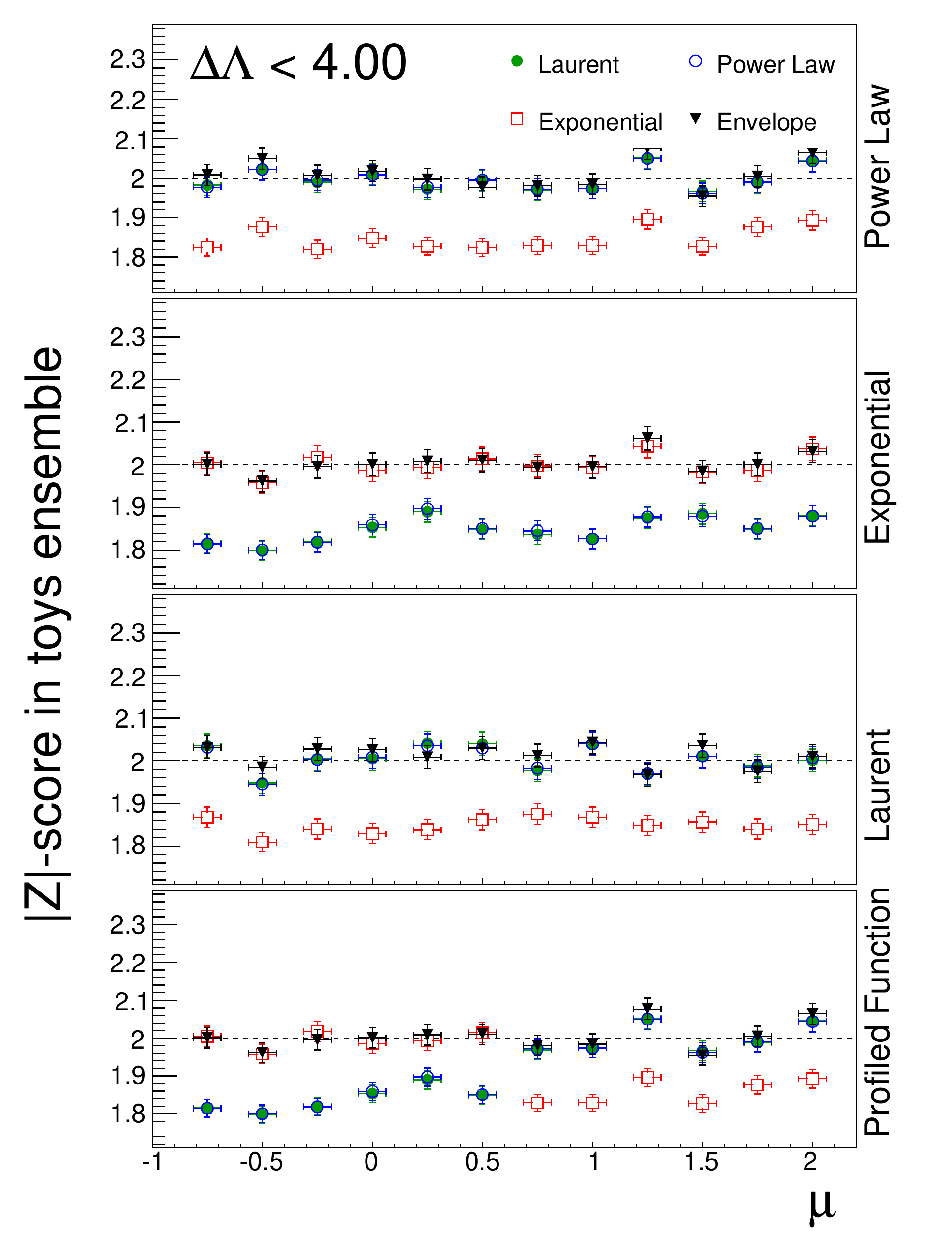}
\label{fig:functions:firstordercoverage:c}}
  \subfigure[]{\includegraphics[width=0.4\textwidth]{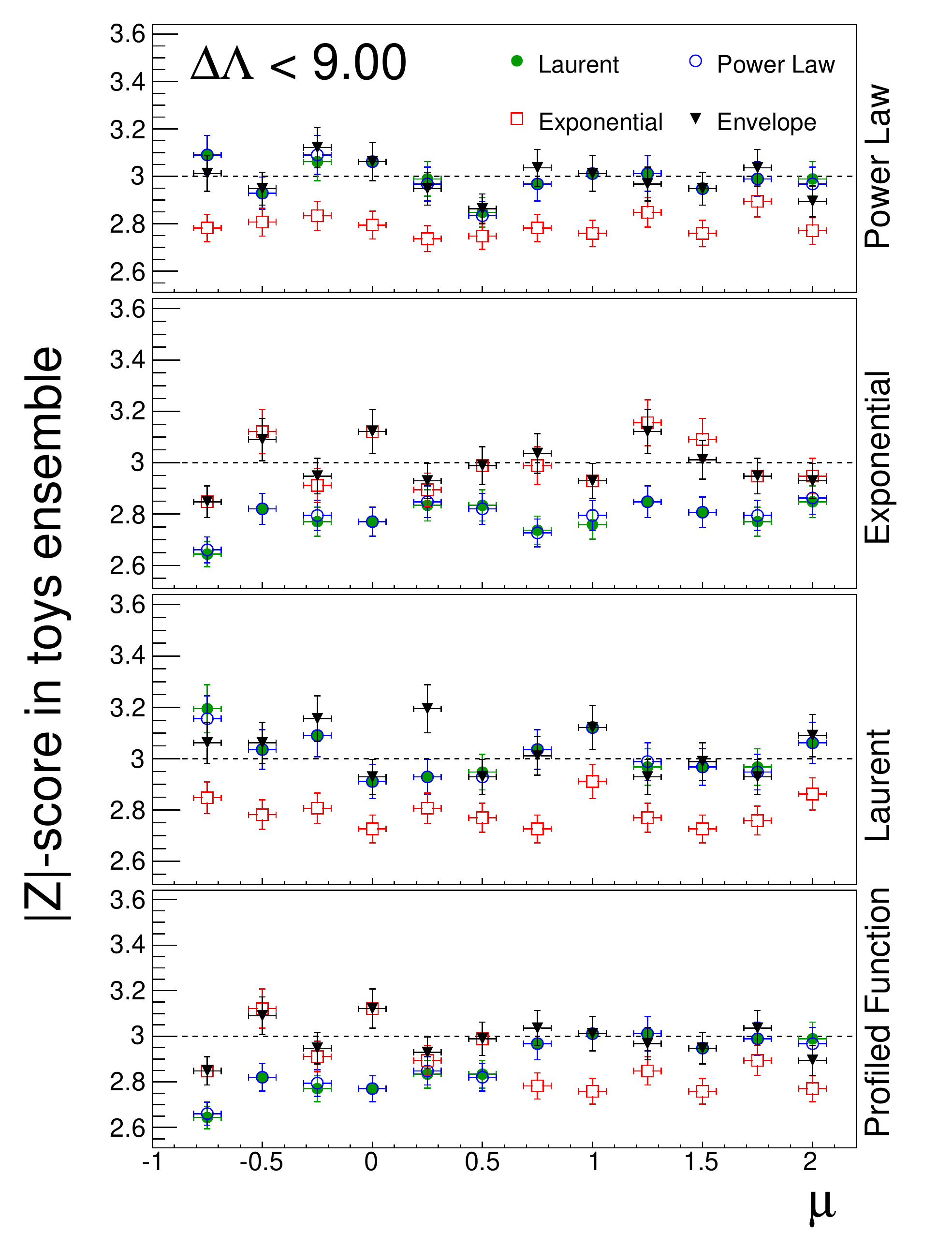}
\label{fig:functions:firstordercoverage:d}}
\caption{Measure of the fraction of the toys which do not contain the generated
$\mu$ in various specified $\Lambda$ intervals, converted
into the two-sided score $|Z|$.
The intervals are
$\Delta\Lambda$ of 0.25 (a), 1.0 (b), 4.0 (c) and 9.0 (d).
These results are obtained by fitting using a single
function and using the envelope.
Within each subfigure,
the first, second and third plots shows the results
when the generating background function is the power law, exponential and Laurent,
respectively. The last plot in each subfigure
shows the result when the best-fit function at each
value of $\mu$ is used to generate toys. Within each panel the different
points correspond to a different fitting function: Laurent (solid green), power law (open blue), exponential (open red) and the envelope of all four two-parameter functions (solid black). In all cases,
fitting with the polynomial gives values outside the range of these plots.}
\label{fig:functions:firstordercoverage}
\end{figure}

The coverage was tested, using the same fits, by determining the frequency with which the difference in \nll between the best fit value $\hat{\mu}$ and the value $\mu$ used to generate the signal was more than 0.25, 1, 4 and 
9~\footnote{These correspond to toys which lie outside the corresponding
standard deviations for a normal distribution.}.
These were converted to the modulus of a two-sided $Z$-score $|Z|$, and
compared with the expected values, namely
0.5, 1, 2 and 3, respectively.
The results for these are shown in figure~\ref{fig:functions:firstordercoverage}. It can be seen that when the fitting function is different from the generating function the calculated confidence interval can undercover, whereas when using the discrete profiling method the coverage is good regardless of the generating function. This good coverage property is found to be
independent of the value of $\mu$ used to generate the signal.








%% file: correction.tex
\section{Using functions with different numbers of parameters} 
\label{sec:correction}

\subsection{Corrections to the likelihood}
\label{sec:correction:corrections}

The \nll calculated from the fit of a function to a dataset is purely a measure
of the agreement of the function and the data; the
number of parameters $N_{\rm par}$
used in the function has no impact. Hence, at least for nested families of
functions (such as polynomials of varying order), then the lowest \nll will
always be given by the highest order considered.
Because this function also has the
largest number of parameters, it will generally have the largest statistical
error and hence widest profiled \nll curve.
Hence, simply using the \nll values
without any ``penalty'' for the number of parameters would effectively
always result in the minimum envelope being mainly defined by the highest
order function considered. It would also mean there is no ``natural'' way to know when
to stop considering yet higher order functions. However, decisions based
on quantities, such as an F-test~\cite{ref:ftest},
are standardly used to determine when higher order functions in
a family can be ignored. Hence, when using functions with different
numbers of parameters, it seems necessary to have some correction to the \nll
value to account for this difference in number.

The idea is therefore to compare \nll values,
correcting for the differing number of parameters
(or equivalently differing number of degrees of freedom) in
the fit functions. Specifically here, where applied, the correction is
done so as to get a value equivalent to a function with no parameters, so
the number of degrees of freedom is the number of bins $N_{\rm bin}$ used.
Two methods were considered, based on the
$\chi^2$ p-value and the Akaike information
criterion~\cite{ref:correction:akaike}.
\begin{enumerate}
\item 
For a binned fit using the expression for the \nll ratio
for each bin specified in equation~\ref{eqn:introduction:def2NLL}, then for
the large statistics case, the \nll becomes equivalent to a $\chi^2$ for the
fit. In this case, it is meaningful to find the p-value of the $\chi^2$ value,
which also depends on the number of degrees of freedom.
A new $\chi^{\prime 2}$
value can now be obtained, namely that which would give the same p-value but
with a different number of degrees of freedom,
equal to the number of bins.
Explicitly, the p-value $p$ is the upper
tail integral of the $\chi^2$ probability
distribution which we write as
\begin{displaymath}
p = F(\chi^2,N_{\mathrm{bin}}-N_{\mathrm{par}})\qquad\mathrm{so}\qquad
\chi^2 = F^{-1}(p,N_{\mathrm{bin}}-N_{\mathrm{par}}),
\end{displaymath}
Hence, the new $\chi^{\prime 2}$ which gives the same p-value with
$N_{\rm bin}$ degrees of freedom is given by
\begin{displaymath}
\chi^{\prime 2} = F^{-1}(p,N_{\rm bin})
\end{displaymath}
and the corrected \nll is then given by
\begin{displaymath}
\Lambda_{\mathrm{corr}} = \textrm{\nll} + (\chi^{\prime 2} - \chi^2).
\end{displaymath}
In the work presented here,
this correction was applied even though some bins in the fits
have lower statistics,
such that the \nll is not a particularly good approximation to the $\chi^2$.

Besides being a function of the number of bins and parameters,
the size of the correction $\chi^{\prime 2} - \chi^2$
depends on the original fit quality
(or equivalently p-value). Figure~\ref{fig:correction:DeltaChiSq}
shows examples of the size of the correction as a function of the
fit p-value, when correcting for various numbers of parameters.
The correction is monotonically decreasing as the p-value gets larger.
Hence, when correcting the profile likelihood curve, the fits further away from
the best fit minimum will get a larger correction. Hence, the profile curve
becomes steeper, which could in principle affect the coverage, even if
only considering one fit function.
However, it can be seen that the correction is approximately given by
\begin{displaymath}
\chi^{\prime 2} - \chi^2 \approx N_{\rm par},
\qquad{\rm so}\qquad
\Lambda_{\mathrm{corr}} \approx \textrm{\nll} + N_{\rm par}
\end{displaymath}
for the central range of p-values. This approximation means the
$\Lambda_{\mathrm{corr}}$ shape, and hence coverage, is unchanged when considering one function alone. Both the
exact p-value
correction and the $N_{\rm par}$ approximate correction are studied in this paper.
\begin{figure}[tbp]
\centering
 \subfigure[]{\includegraphics[width=0.46\textwidth]{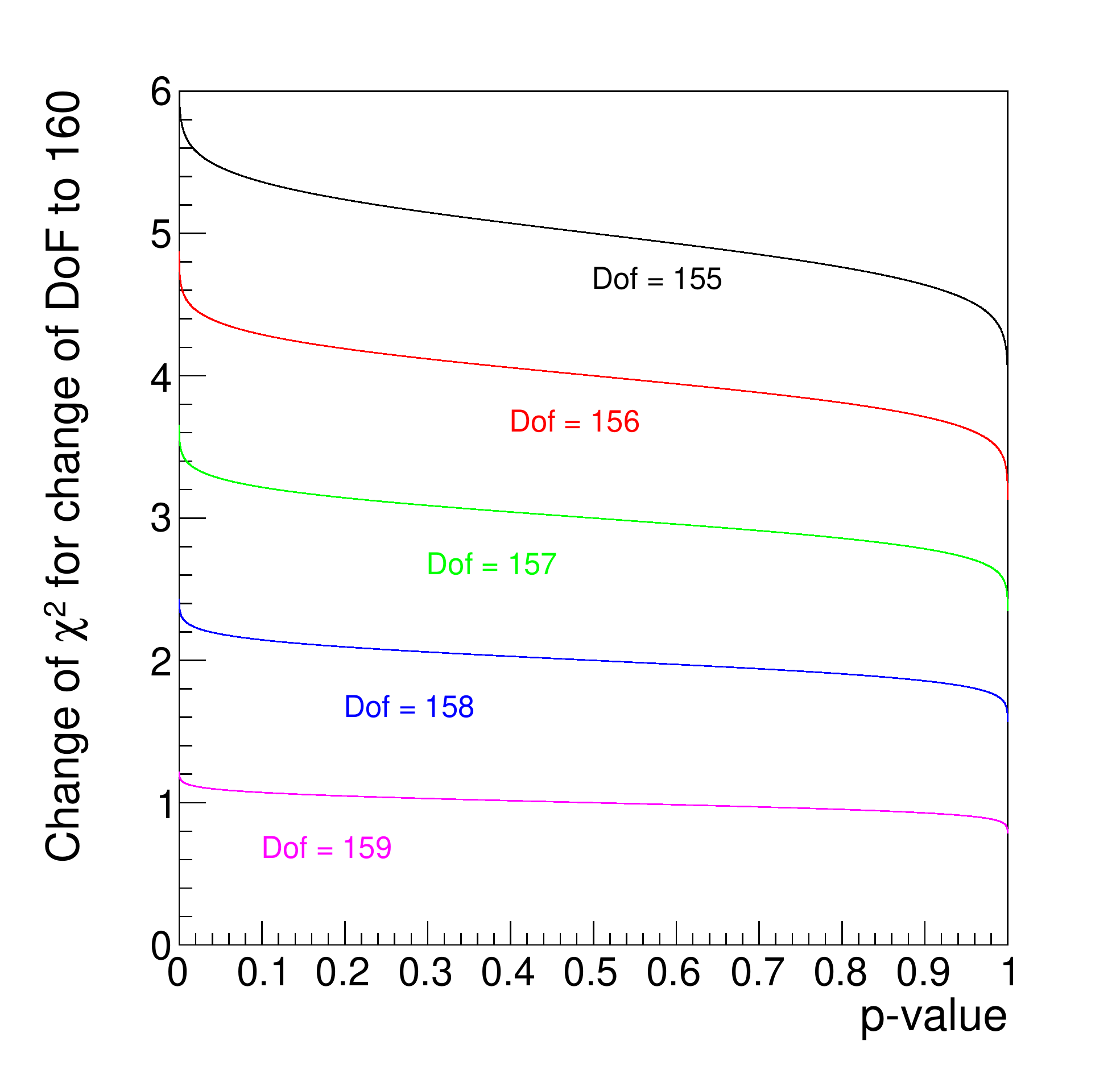}
\label{fig:correction:DeltaChiSq:a}}
 \subfigure[]{\includegraphics[width=0.46\textwidth]{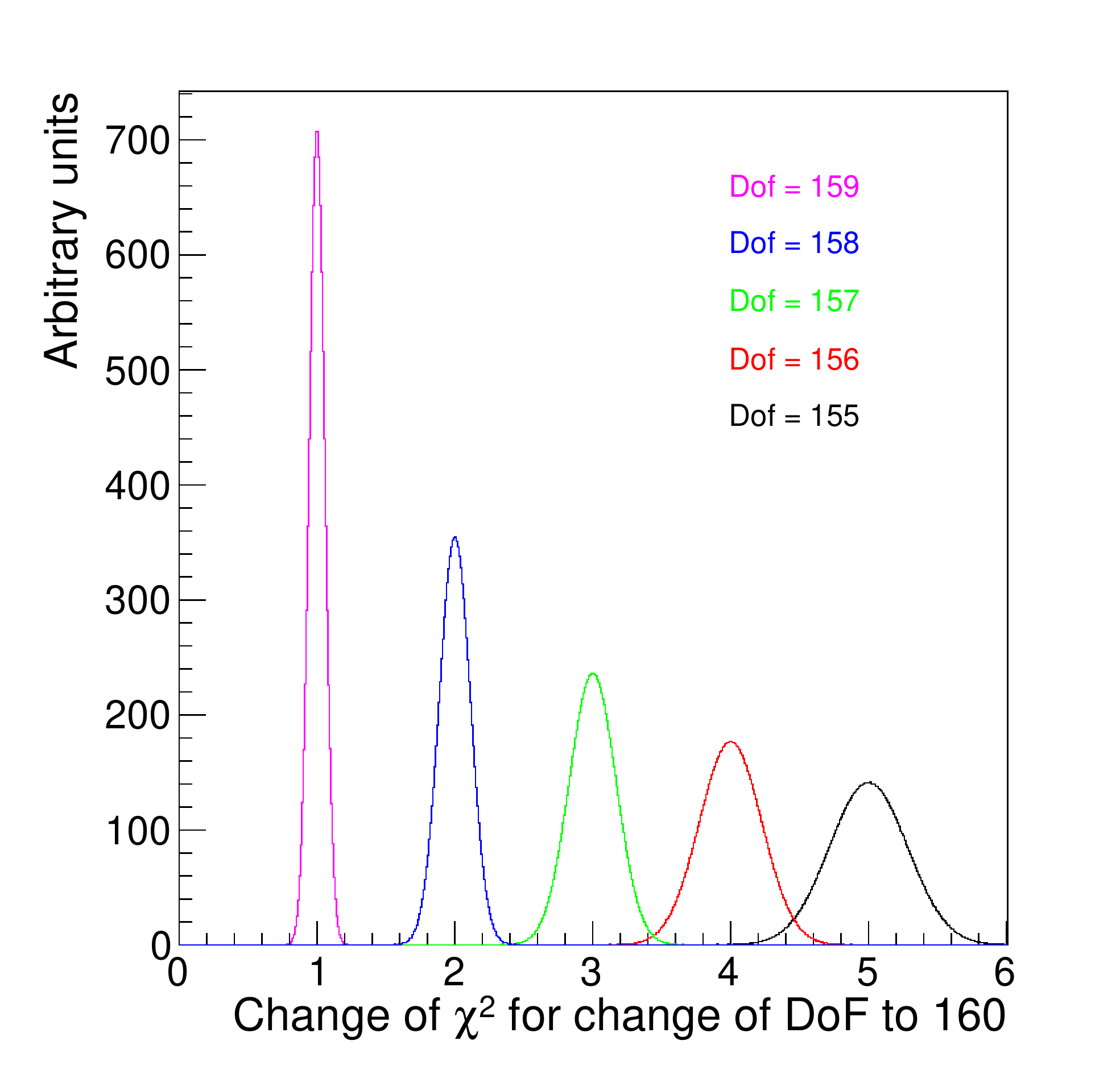}
\label{fig:correction:DeltaChiSq:b}}
\caption{Change of $\chi^2$ when correcting for between one and five parameters
in a fit with 160 bins. (a) The change to the $\chi^2$ as a function of the
original p-value. (b) The distribution of the change of
the $\chi^2$ assuming a flat p-value distribution.}
\label{fig:correction:DeltaChiSq}
\end{figure}

\item 
The basic Akaike formula for very large sample sizes is
\begin{displaymath}
\Lambda_{\mathrm{corr}} = \textrm{\nll} + 2N_{\rm par},
\end{displaymath}
so the correction is simply $ 2N_{\rm par}$ and
hence is twice as big as the approximate p-value correction.
For finite samples, then it is modified to
\begin{displaymath}
\Lambda_{\mathrm{corr}}
= \textrm{\nll} + 2N_{\rm par}  + \frac{2N_{\rm par}(N_{\rm par}+1)}{n-(N_{\rm par}+1)}
= \textrm{\nll} + \frac{2N_{\rm par}}{1-(N_{\rm par}+1)/n},
\end{displaymath}
where $n$ is the sample size, i.e.~the number of bins (for a binned likelihood
fit) or events (for an unbinned likelihood fit). For large $n$, the original
formula is clearly regained.
Also the correction does not depend on the \nll value and
so is a simple shift of
the whole profile curve, without changing its shape.
Hence, it also has no effect on coverage when considering one function alone.
\end{enumerate}

\subsection{Function definitions}
\label{sec:correction:functions}
The fit functions which were used are the two-parameter functions listed in
Section~\ref{sec:functions:function} and higher order generalisations of
these. Specifically, these are
\begin{enumerate}
\item
``Power law sum''; $f(x) = \sum_{i=0}^N p_{2i} x^{p_{2i+1}} = p_{0}x^{p_{1}} + p_{2}x^{p_{3}} + p_{4}x^{p_{5}}+\cdots$.
\item
``Exponential sum''; $f(x) = \sum_{i=0}^N p_{2i} e^{p_{2i+1}x} = p_{0} e^{p_{1}x} + p_{2} e^{p_{3}x} + p_{4} e^{p_{5}x}+\cdots$.
\item
``Laurent series''; $f(x) = \sum_{i=0}^N p_i/x^{n_i}$.
\item
``Polynomial''; $f(x) = \sum_{i=0}^N p_i x^i$.
\end{enumerate}
The Laurent function values of $n_{i}$ used for $i=0,1,2,3,4,5\dots$ are
$n_{i}=4,5,3,6,2,7\dots$, meaning they are grouped around the original
two $n_{i}$ values of 4 and 5 as used throughout Section~\ref{sec:functions}.
Note the power law and exponential functions can only have even numbers of
parameters, i.e.~$N_{\rm par}=2N$, while the Laurent and polynomial functions
can have both even and odd numbers, i.e.~$N_{\rm par}=N$.

\subsection{Example case}
\label{sec:correction:example}

\begin{figure}[tbp]
\centering
\subfigure[]{
 \includegraphics[width=0.46\textwidth]{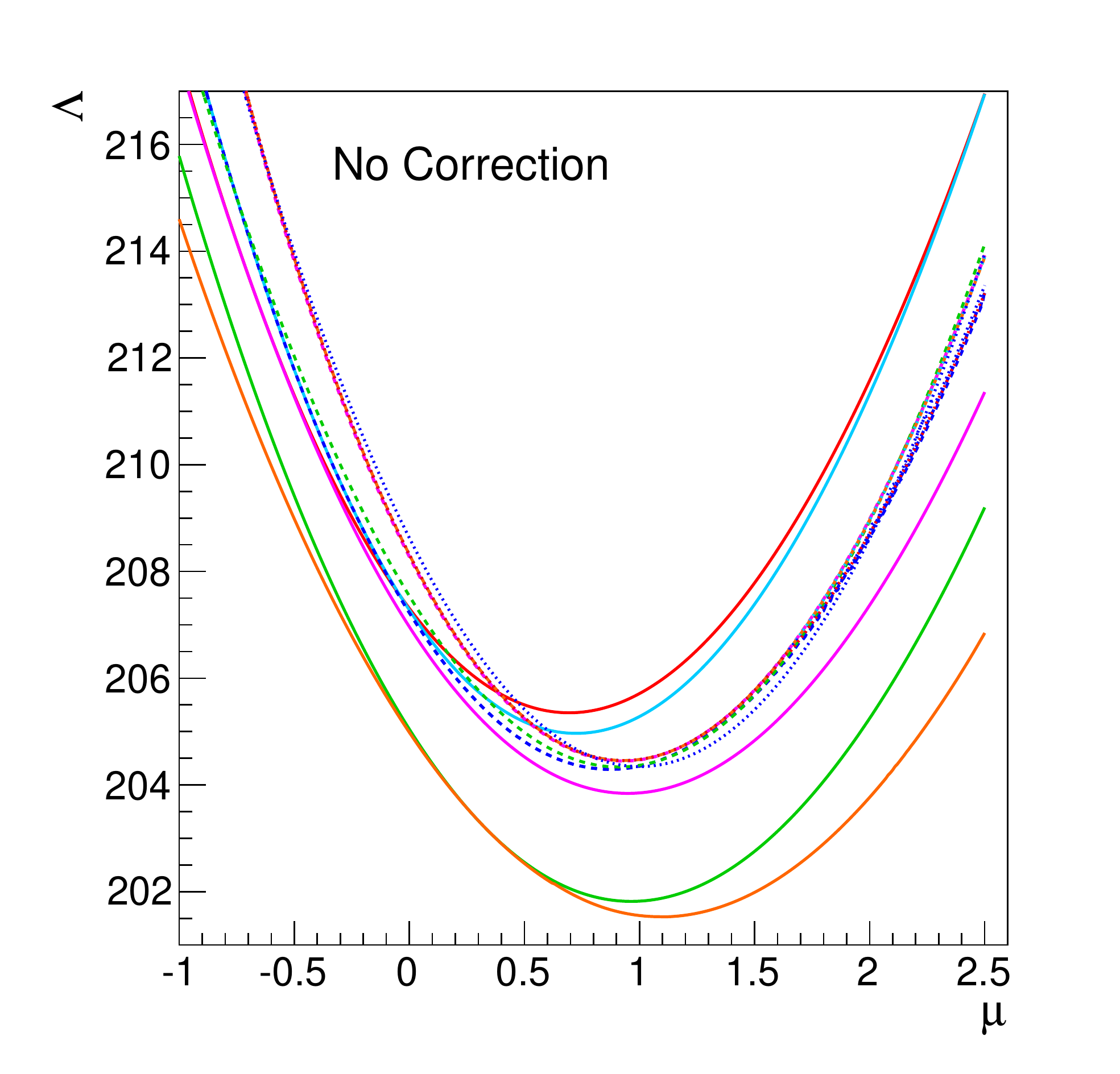}
 \includegraphics[width=0.46\textwidth]{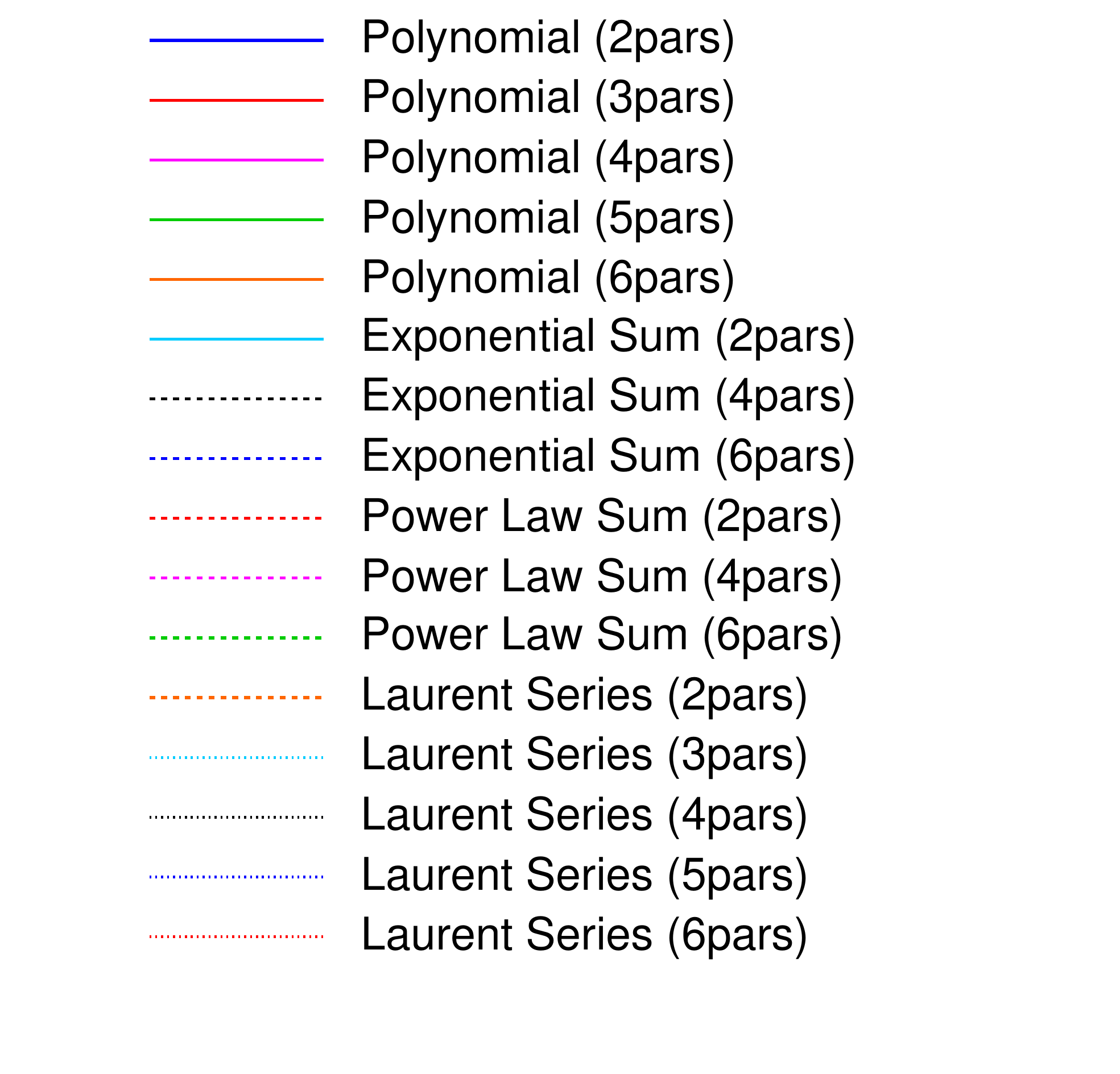}
\label{fig:correction:profiles-no:a}
}
\subfigure[]{
 \includegraphics[width=0.46\textwidth]{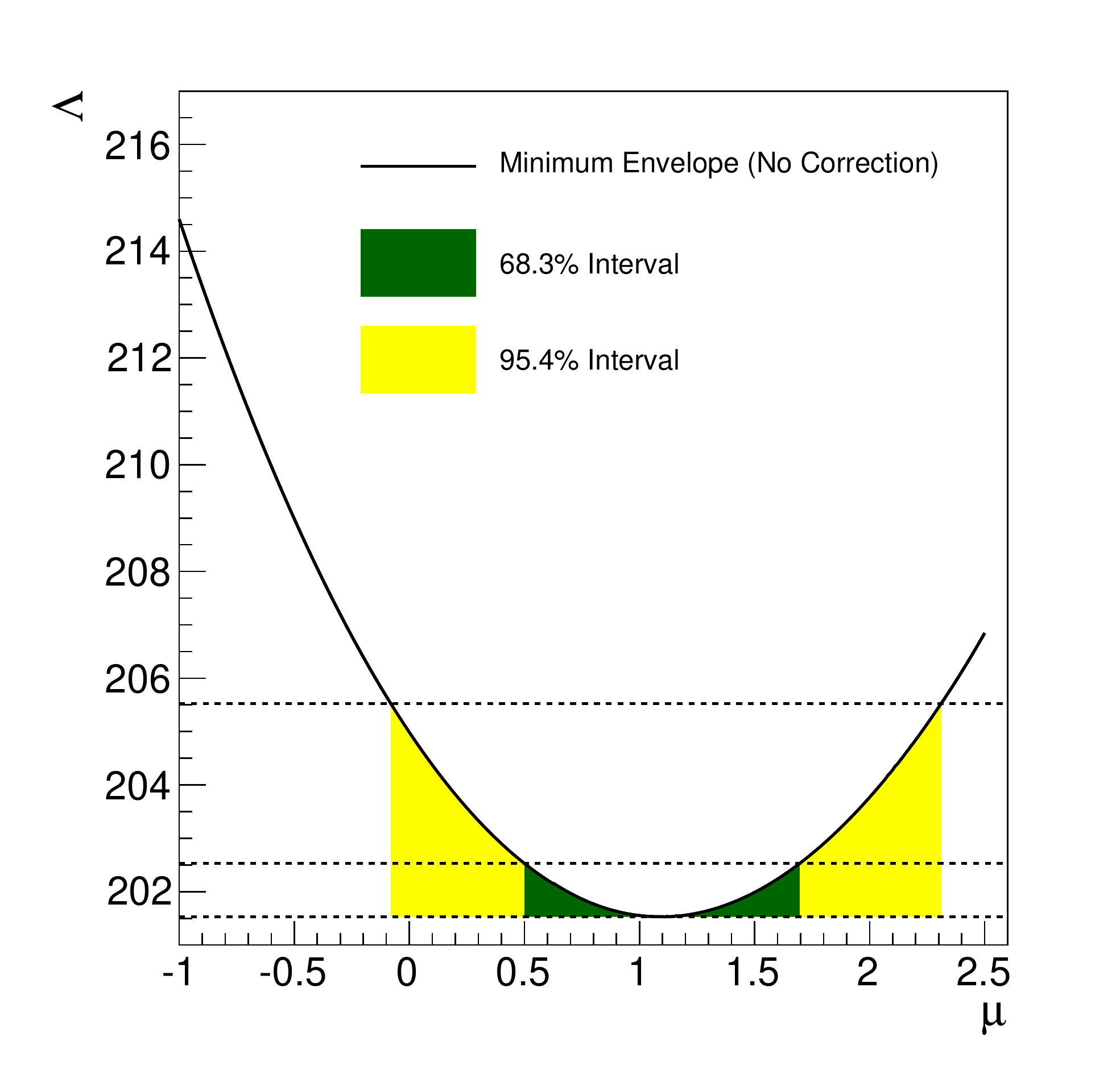}
\label{fig:correction:profiles-no:b}
}
\caption{(a) Profiled \nll curves for all of the functions considered. The \nll value for each profile curve has not been corrected. The labels indicate the function and the value of $N_{\rm par}$. (b) Minimum envelope of the functions. The \nll scan when only considering the best fit function is drawn in red, although this is completely obscured by the envelope curve in this case.}
\label{fig:correction:profiles-no}
\end{figure}

\begin{figure}[tbp]
\centering
\subfigure[]{
 \includegraphics[width=0.46\textwidth]{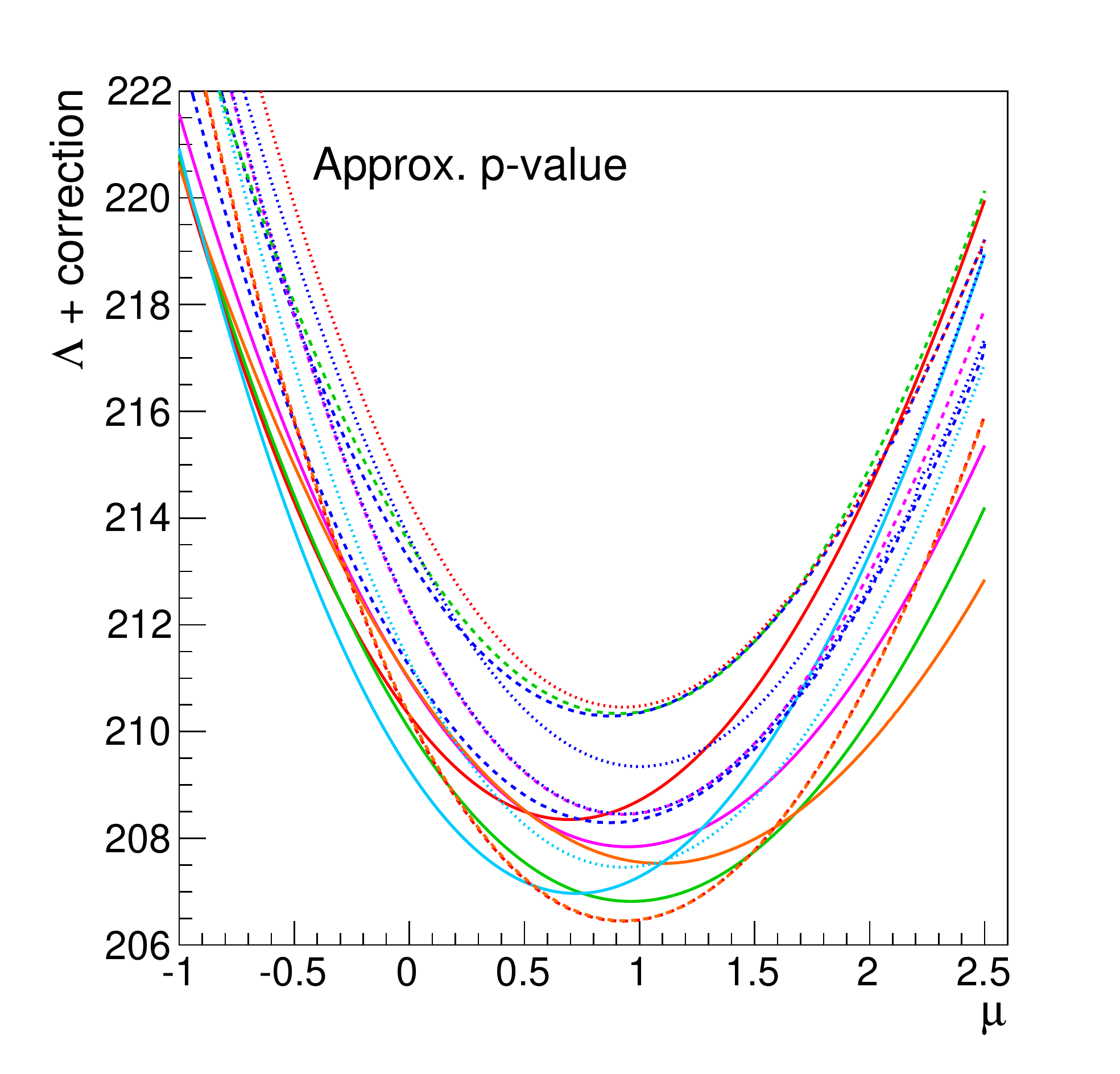}
 \includegraphics[width=0.46\textwidth]{{correction/AllFunctionsLegend}.pdf}
\label{fig:correction:profiles-pval:a}
}
\subfigure[]{
 \includegraphics[width=0.46\textwidth]{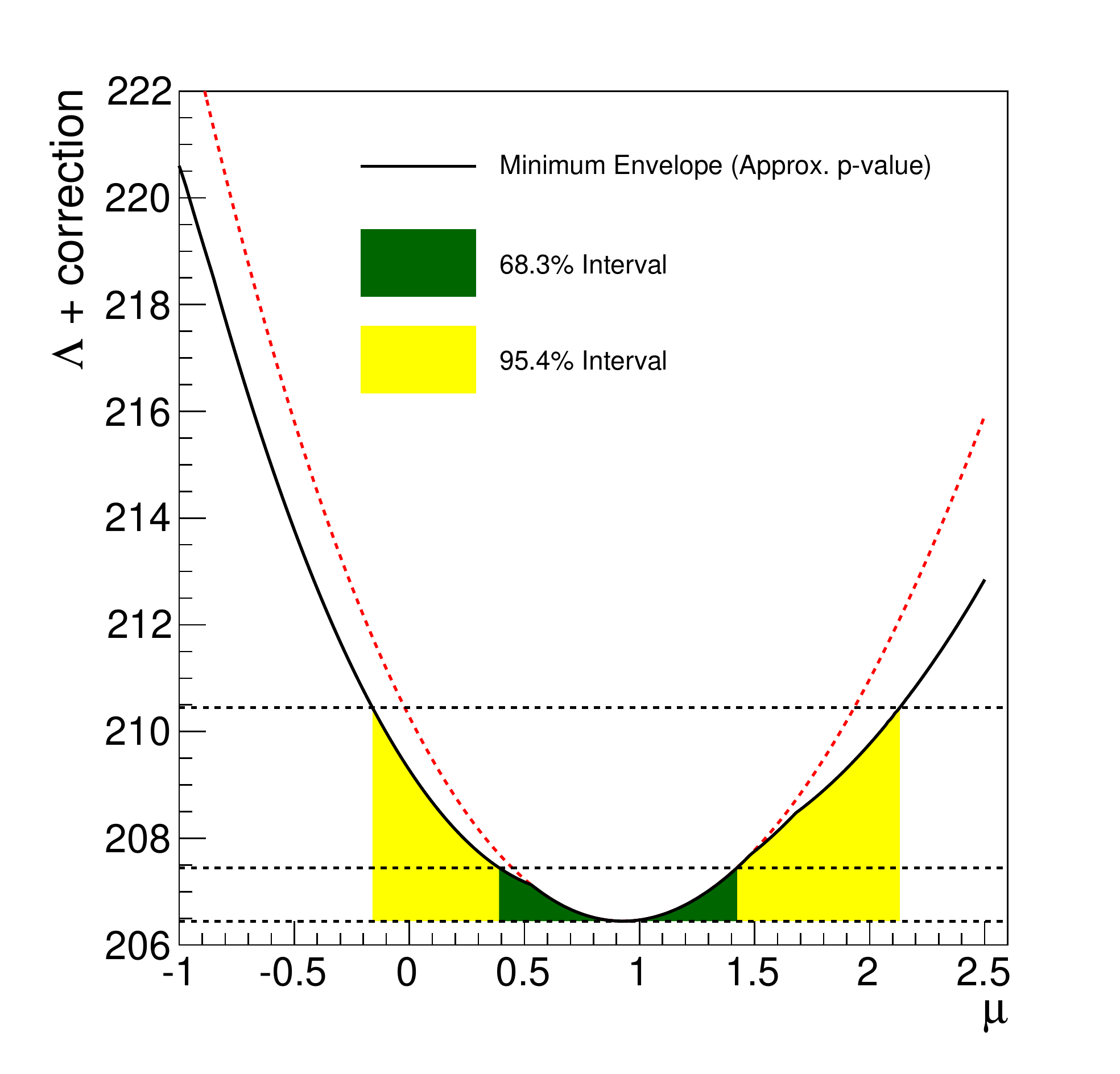}
\label{fig:correction:profiles-pval:b}
}
\caption{(a) Profiled \nll curves for all of the functions considered. The \nll value for each profile curve has been corrected using the approximate p-value correction of 1 per background parameter. The labels indicate the function and the value of $N_{\rm par}$. (b) Minimum envelope of the functions after applying a correction of 1 per background parameter to each \nll curve. The \nll scan when only considering the best fit function is shown in red.}
\label{fig:correction:profiles-pval}
\end{figure}

\begin{figure}[tbp]
\centering
\subfigure[]{
 \includegraphics[width=0.46\textwidth]{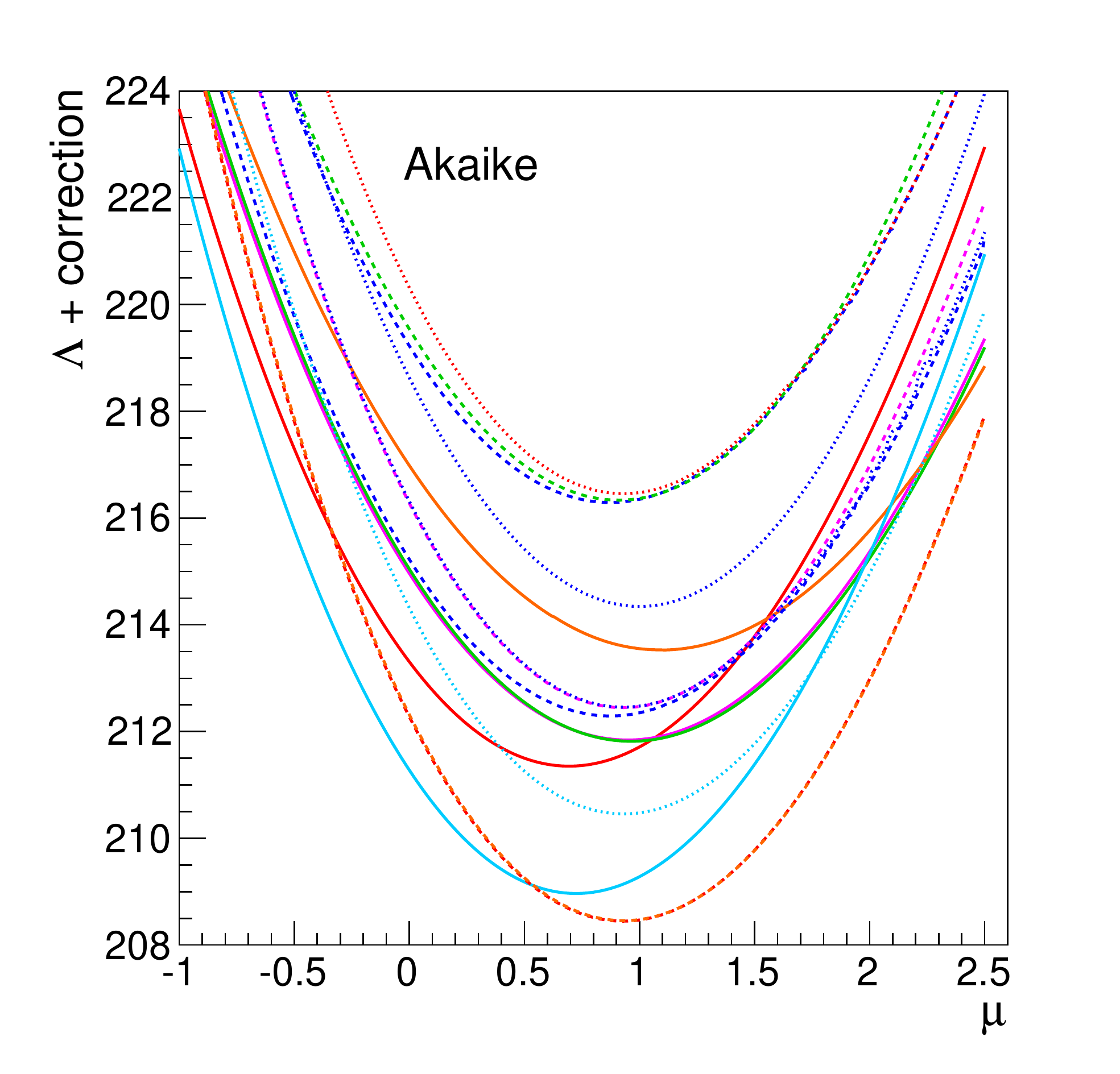}
 \includegraphics[width=0.46\textwidth]{{correction/AllFunctionsLegend}.pdf}
\label{fig:correction:profiles-akaike:a}
}
\subfigure[]{
 \includegraphics[width=0.46\textwidth]{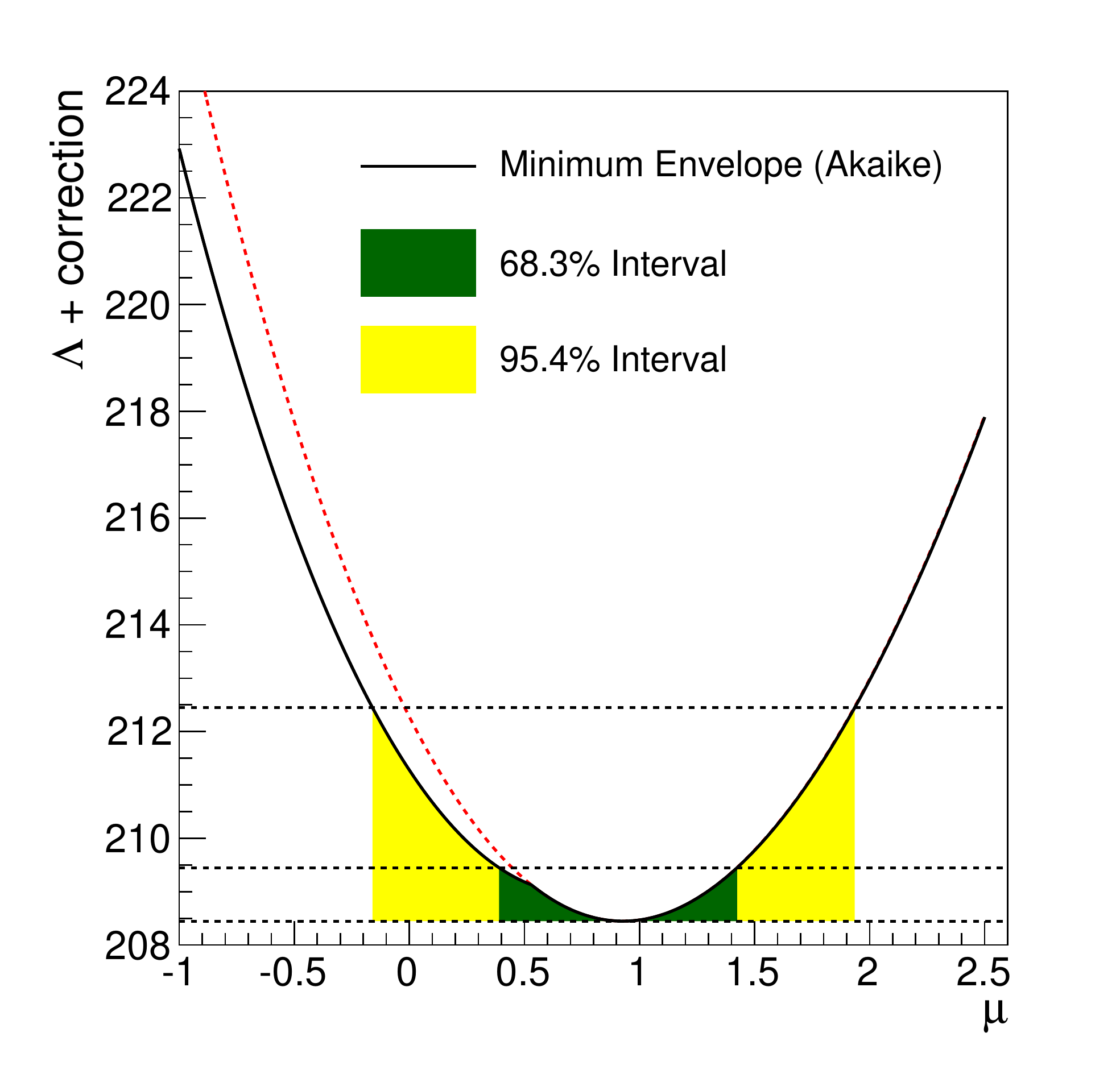}
\label{fig:correction:profiles-akaike:b}
}
\caption{(a) Profiled \nll curves for all of the functions considered. The \nll value for each profile curve has been corrected using the Akaike correction of 2 per background parameter. The labels indicate the function and the value of $N_{\rm par}$. (b) Minimum envelope of the functions after applying a correction of 2 per background parameter to each \nll curve. The \nll scan when only considering the best fit function is shown in red.}
\label{fig:correction:profiles-akaike}
\end{figure}

The functions listed above were fit to the original dataset for values of
$2 \le N_{\rm par} \le 6$; this resulted in three fits for the power law and
exponential functions, and five fits for the Laurent and polynomial functions.
This range was chosen for practical purposes as these functions gave reasonable
fits, without requiring larger numbers of parameters.
The profile curves from the fits of these functions are shown in
figures~\ref{fig:correction:profiles-no},~\ref{fig:correction:profiles-pval} and~\ref{fig:correction:profiles-akaike} where results for three different
corrections to \nll have been shown, namely no correction, the approximate
p-value correction, and the Akaike correction respectively.
These figures also show the 68.3\% and
95.4\% confidence intervals which would be determined from the
profile envelope. 

Consider the example case of the approximate p-value correction,
i.e.~correcting by one unit per background function parameter,
which is shown
in figure~\ref{fig:correction:profiles-pval}.
For this case, the lowest corrected \nll value is still given by the
two-parameter power law function and so gives an identical central value
to that described in Section~\ref{sec:functions:example}. The lowest corrected
\nll value for $\mu < 0.55$ is also again given by the two-parameter exponential
function, also as for the previous case.
Indeed, comparison with figure~\ref{fig:functions:profiles} shows that the
minimum values of the corrected \nll for low $\mu$
are simply two units larger here than in that figure.
However, for $1.48 < \mu < 1.68$,
the $N_{\rm par}=5$ polynomial is the lowest function in the profile,
while for $\mu > 1.68$,
the $N_{\rm par}=6$ polynomial is lowest.
Hence, the envelope is formed from four different functions in this
case. 
The 68.3\% region is identical to that found using
just the two-parameter functions (see Section~\ref{sec:functions:example}), but
when including higher order functions, the 95.4\% confidence interval
on $\mu$ is $-0.18 < \mu < 2.11$,
i.e.~it is extended to higher $\mu$
due to the influence of the higher order polynomial functions providing a
reasonable description of the data.
%

In contrast, the case with no correction to \nll, shown in
figure~\ref{fig:correction:profiles-no}, shows that the highest order polynomial
function used, with $N_{\rm par}=6$, defines the envelope across the whole
range of $\mu$. This domination of the functions with the highest numbers of
parameters is exactly what the penalty correction to \nll is attempting to
avoid. Finally, figure~\ref{fig:correction:profiles-akaike},
corresponding to the Akaike correction, shows that all functions with more
than the minimum of two parameters get a large penalty and so do not
contribute to the envelope. This hints that this correction may be too severe.

\begin{figure}[htp]
\centering
\includegraphics[angle=90,width=0.38\textwidth]{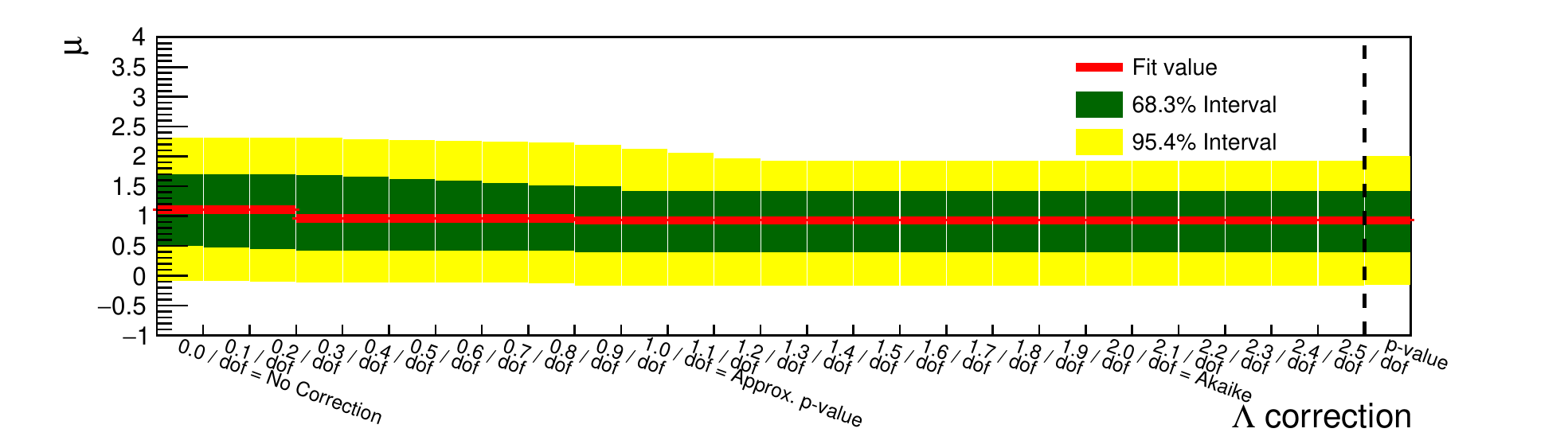}
\caption{The best fit value of the signal strength, $\mu$, and its uncertainty when fitted to the dataset as a function of the correction applied per background parameter.}
\label{fig:correction:correction}
\end{figure}

The three corrections shown can be
considered to be examples of a continuous spectrum of corrections which can
be written as
\begin{displaymath}
\Lambda_{\mathrm{corr}} = \textrm{\nll} + cN_{\rm par},
\end{displaymath}
where $c=0$, 1 or 2 in the cases shown. Other values of $c$ are clearly
also possible.
Figure~\ref{fig:correction:correction} shows the
68.3\% and 95.4\% intervals which would be derived from the envelopes
for these and
some other values of $c$, as well as for the exact p-value correction.
It is seen that the best fit value and intervals change for $c<0.5$ but
for larger values of $c$ they are quite stable. Hence, the result which would
be derived from this method is effectively insensitive to the exact correction
used, as long as $c \gtrsim 1$.




\subsection{Bias and coverage dependence on correction}
\label{sec:correction:bias}

In a similar way to Section~\ref{sec:functions:coverage}, toy datasets
were generated using various individual functions and also the best fit
functions for each $\mu$ value.

\begin{figure}[tbp]
\centering
 \subfigure[]{\includegraphics[width=0.32\textwidth]{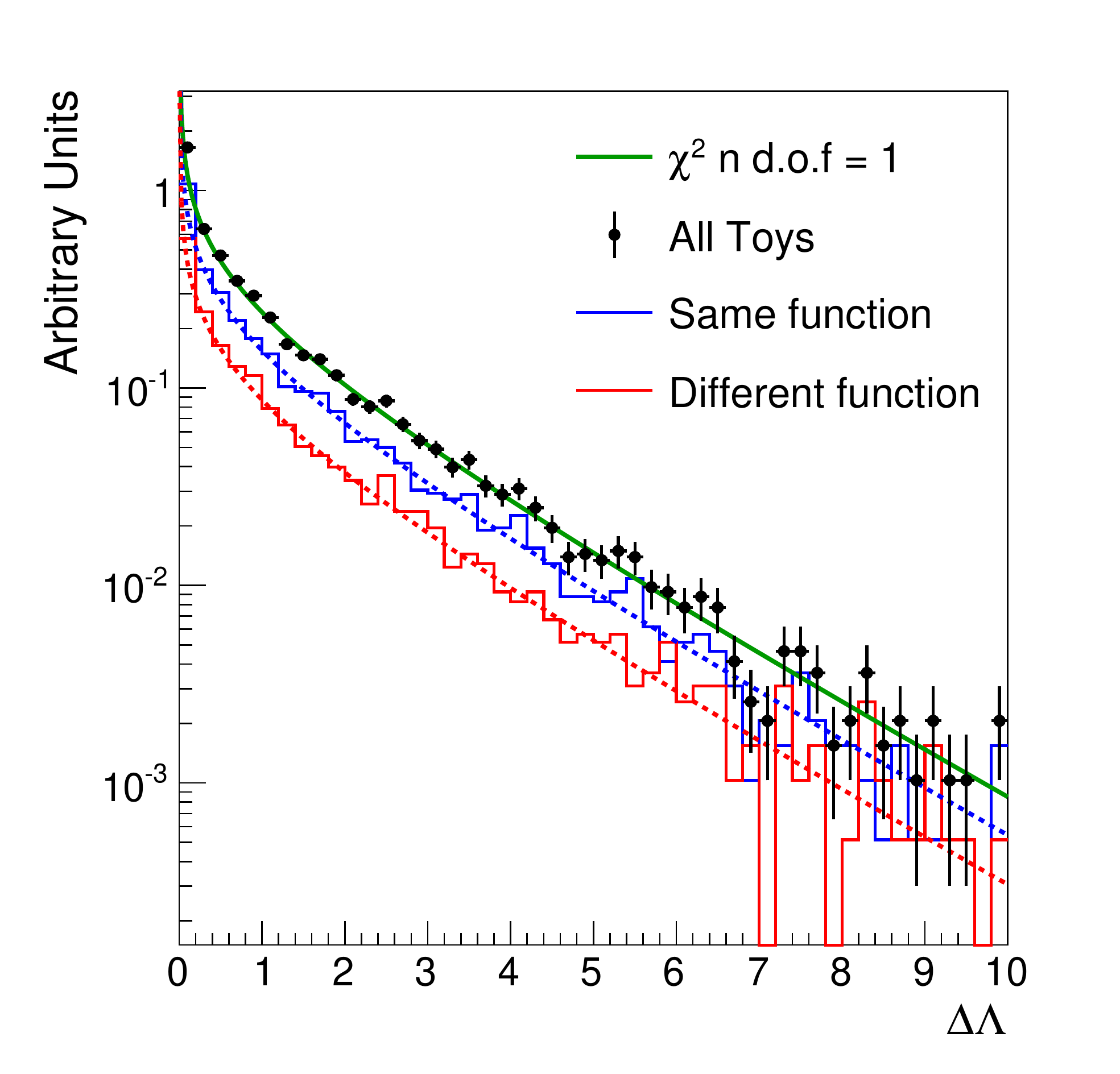}
\label{fig:correction:chiSq:a}}
 \subfigure[]{\includegraphics[width=0.32\textwidth]{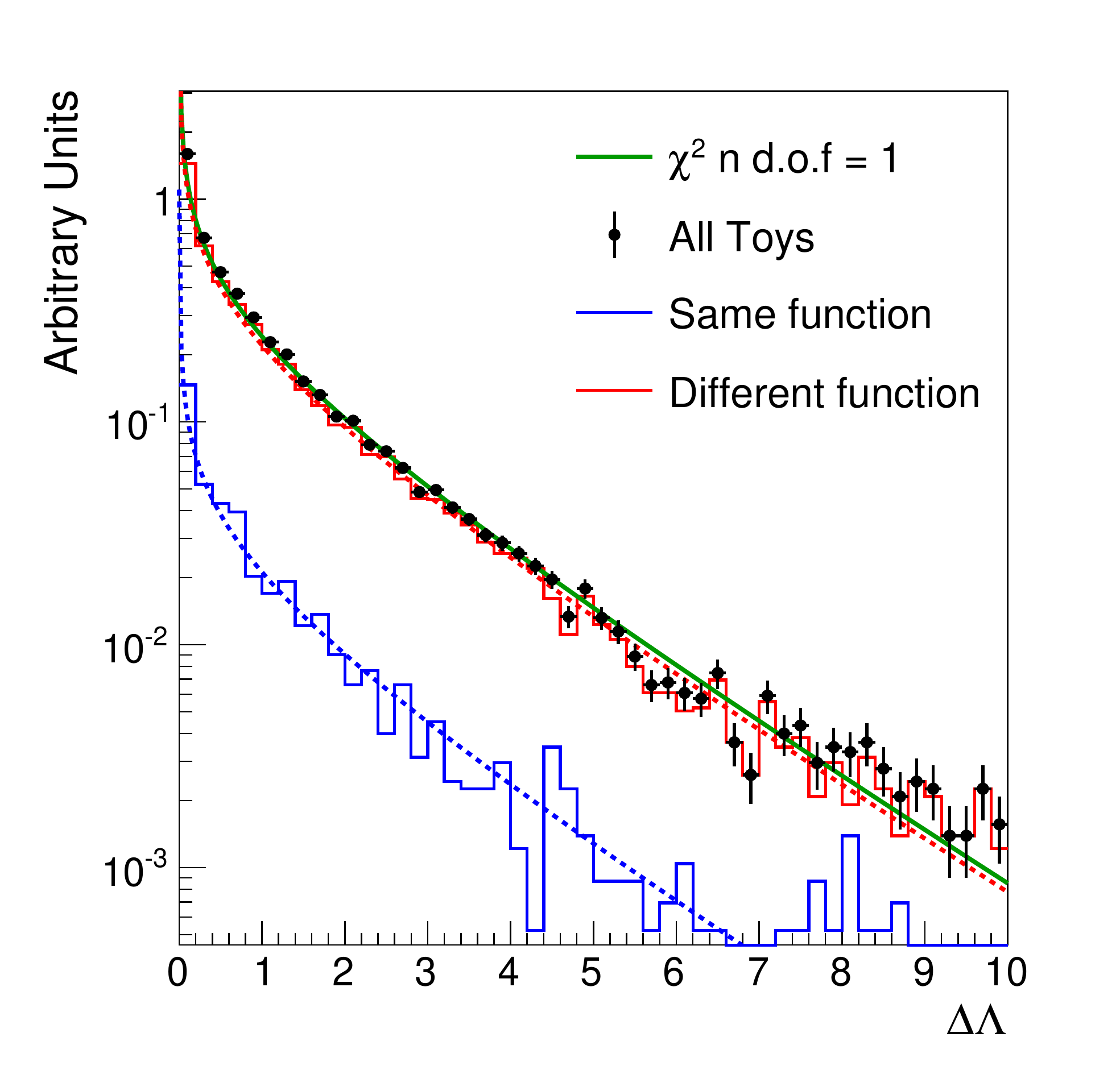}
\label{fig:correction:chiSq:b}}
 \subfigure[]{\includegraphics[width=0.32\textwidth]{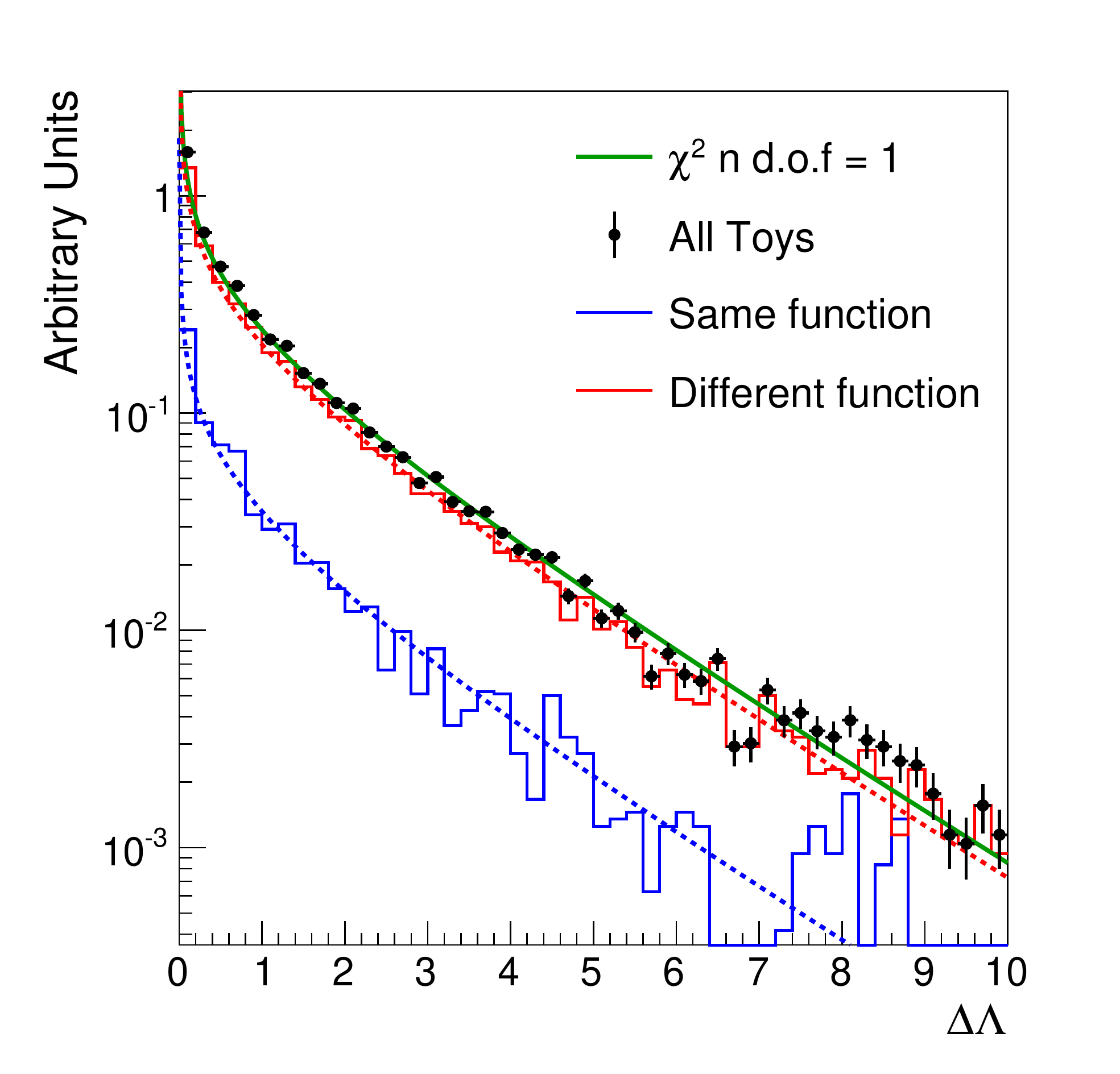}
\label{fig:correction:chiSq:c}}
\caption{Distribution of  $\Delta\nll$, the difference
between the \nll value with $\mu$ fixed to its true value and \nll at the best fit value
of $\mu$, (a) with no correction to \nll, (b) with the approximate p-value
correction, and (c) with the Akaike correction.
The black data points are from the toy dataset fits and the green function
shows the expected $\chi^2$ distribution for one degree of freedom.
The blue and red histograms show the toy values separated into
cases where the best fit uses the same or different functions, respectively, compared with the function used to generate the toys.}
\label{fig:correction:chisq}
\end{figure}

Figure~\ref{fig:correction:chisq} shows examples of
the distribution of the difference
in \nll between the true and best fit values of $\mu$. These are shown for the
three correction methods considered. In each case, the function used to generate
the toy datasets was the one giving the best fit as determined from the 
envelopes in figures~\ref{fig:correction:profiles-no}, 
\ref{fig:correction:profiles-pval} and
\ref{fig:correction:profiles-akaike}.
Similarly to figure~\ref{fig:functions:chisq}, these indicate that the
change in \nll is similar to the expected $\chi^2$ distribution and so
is a reasonable basis on which to estimate the uncertainty using an
asymptotic approximation.

\begin{figure}[tbp]
\centering
\includegraphics[width=0.8\textwidth]{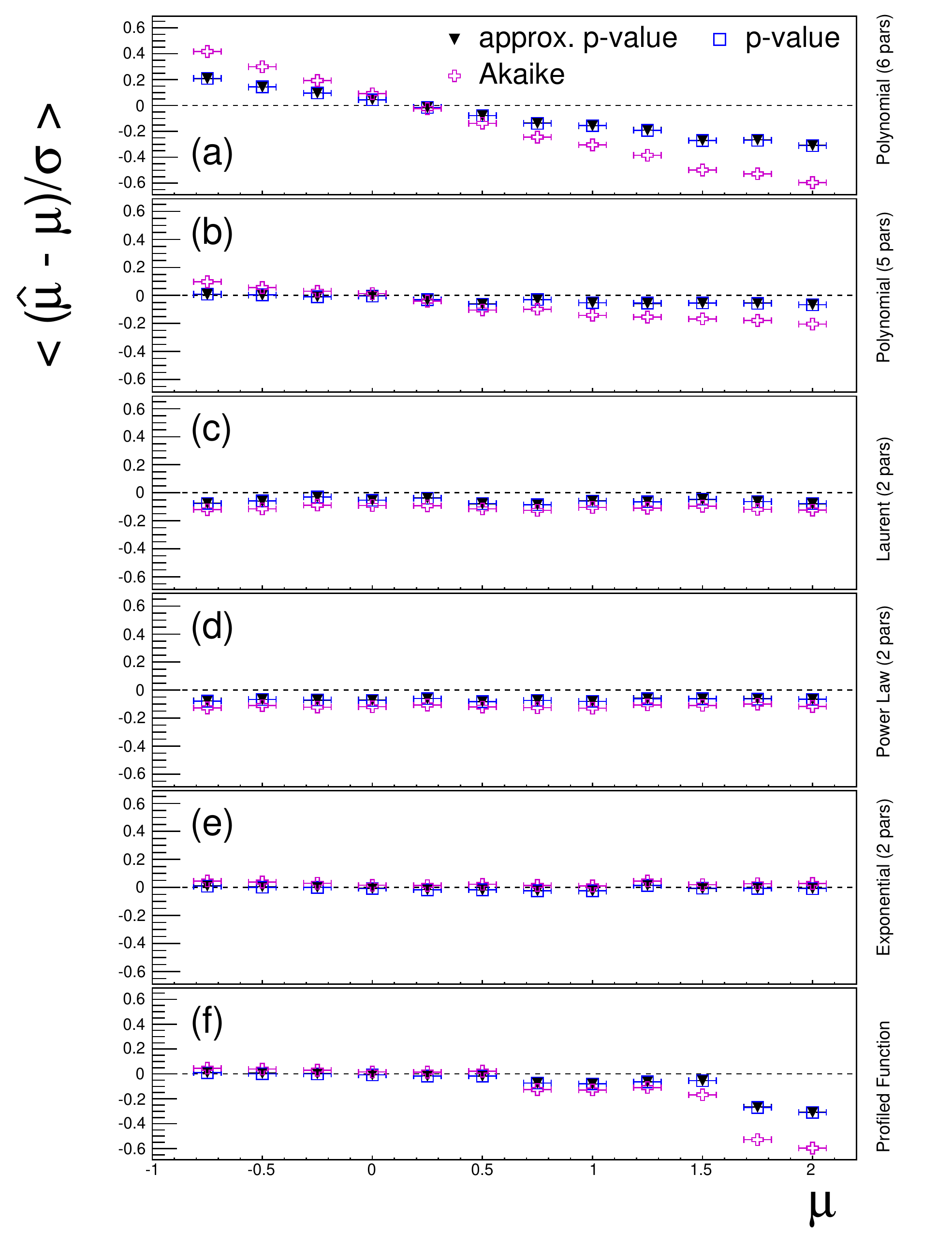}
\caption{Average pull when fitting using the envelope as a function of $\mu$
used to generate the signal when correcting
with the different \nll correction schemes.
Each panel shows the bias when using
a different background function for toy generation, with the functions
used being: polynomial with $N_{\rm par}=6$ (a),
polynomial with $N_{\rm par}=5$ (b), Laurent with $N_{\rm par}=2$ (c),
power law with $N_{\rm par}=2$ (d), and exponential with $N_{\rm par}=2$ (e).
Panel (f) shows the result when the best-fit function at each
value of $\mu$ is used to generate toys.
}
\label{fig:correction:allorderbias}
\end{figure}

Figure~\ref{fig:correction:allorderbias} shows the average pull vs $\mu$
for each choice of generating function, with the envelope based on the
three correction methods considered, i.e.~approximate p-value, p-value
and Akaike.
The generating functions chosen were those
which gave the lowest corrected \nll values in
figure~\ref{fig:correction:profiles-pval}.
Some obvious biases are seen, in particular for the cases when
generating with polynomials of five or six parameters. It seems that in these
cases, which are among the functions with the highest numbers of parameters,
the generated data are too complex to be described well by the lower order
functions. The choice of the Akaike correction makes the bias worse, as it
tends to emphasise the contribution of lower order functions to the envelope.

However,
even with the bias being a significant fraction of the error for some functions,
it is seen that the approximate p-value and p-value corrections give
effectively identical results. It is also found
that the p-value corrections always give less bias than the Akaike correction.

\begin{figure}[tbp]
\centering
\subfigure[]{
\includegraphics[width=0.40\textwidth]{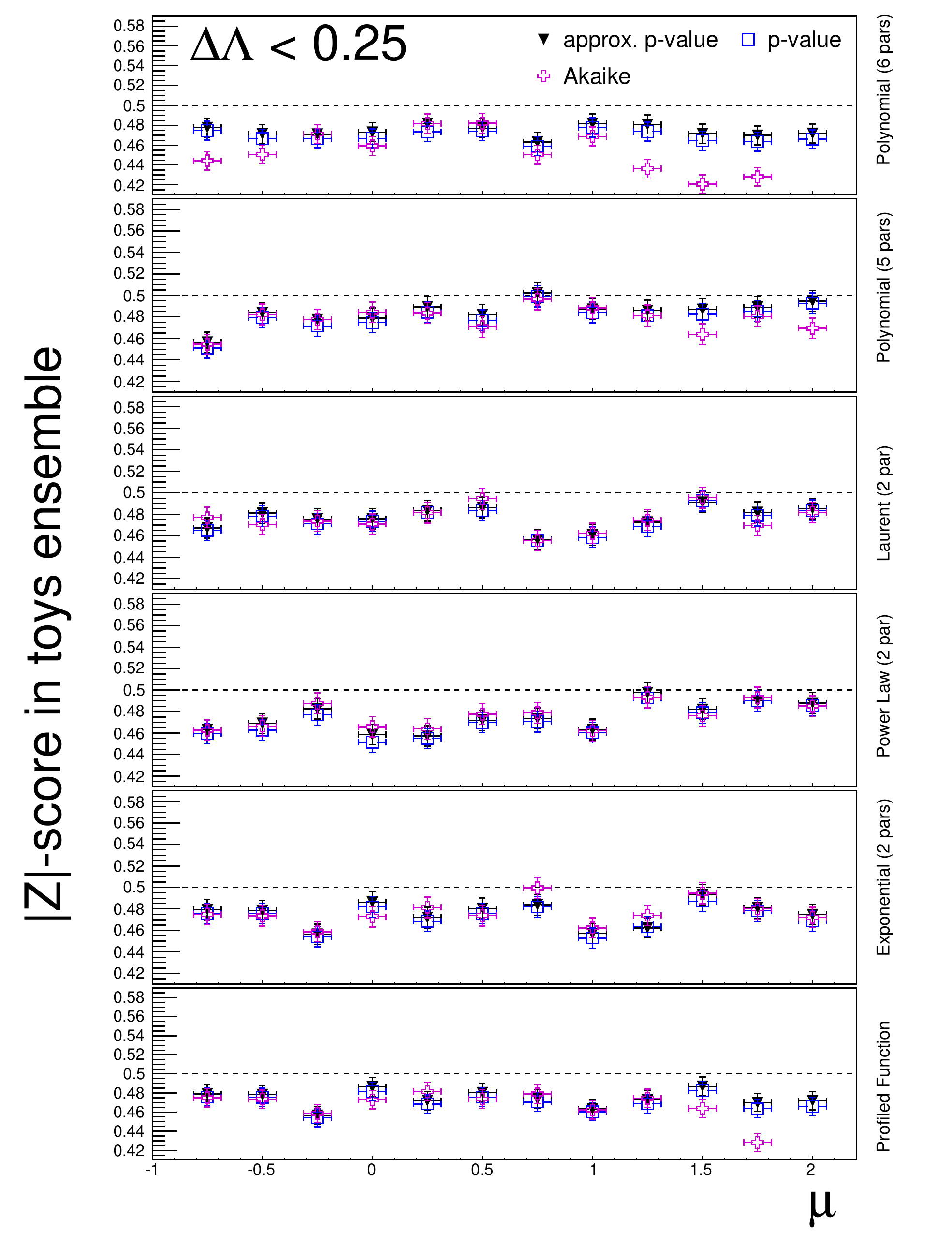}
\label{fig:correction:allorderecoverage:a}
}
\subfigure[]{
\includegraphics[width=0.40\textwidth]{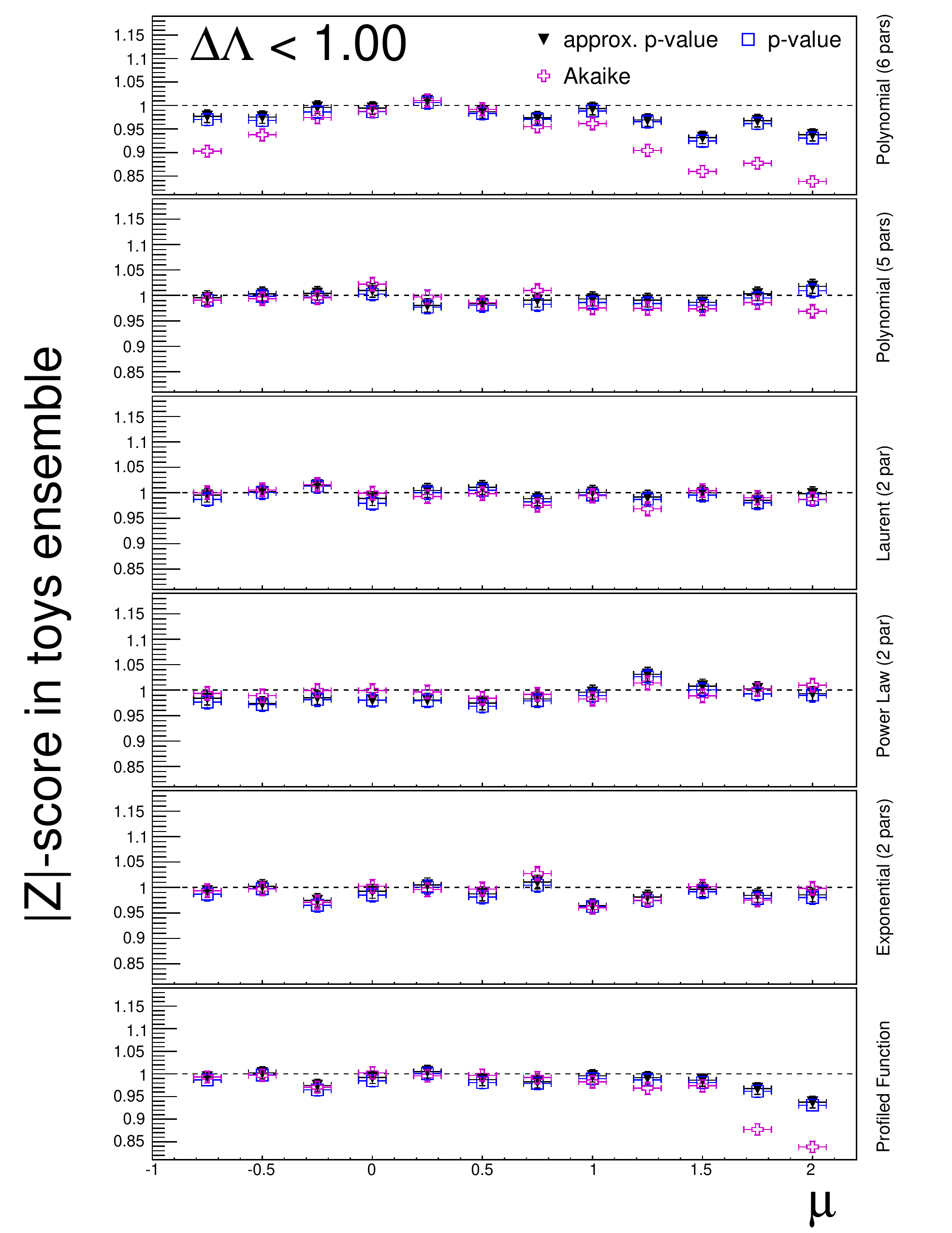}
\label{fig:correction:allorderecoverage:b}
}\\
\subfigure[]{
\includegraphics[width=0.40\textwidth]{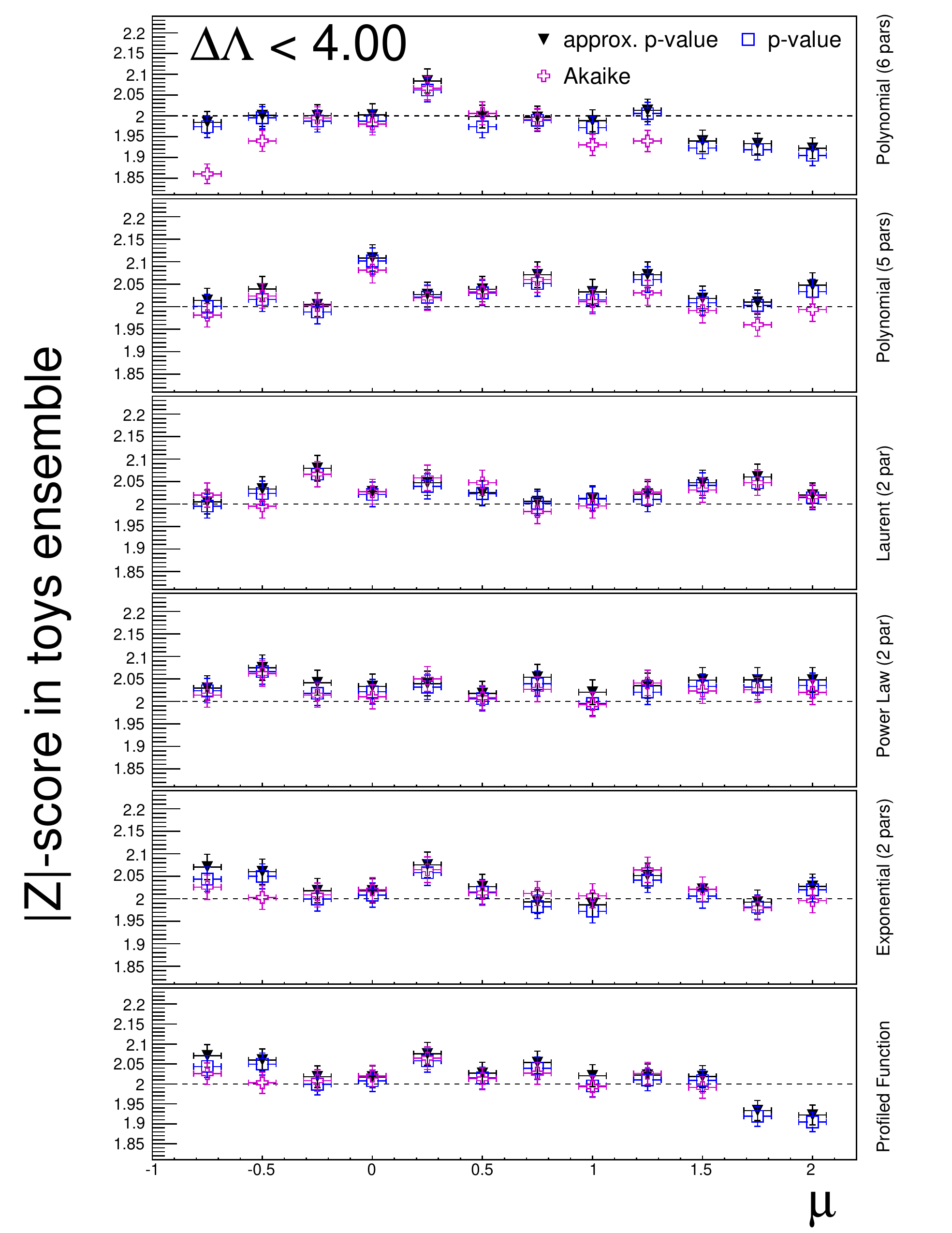}
\label{fig:correction:allorderecoverage:c}
}
\subfigure[]{
\includegraphics[width=0.40\textwidth]{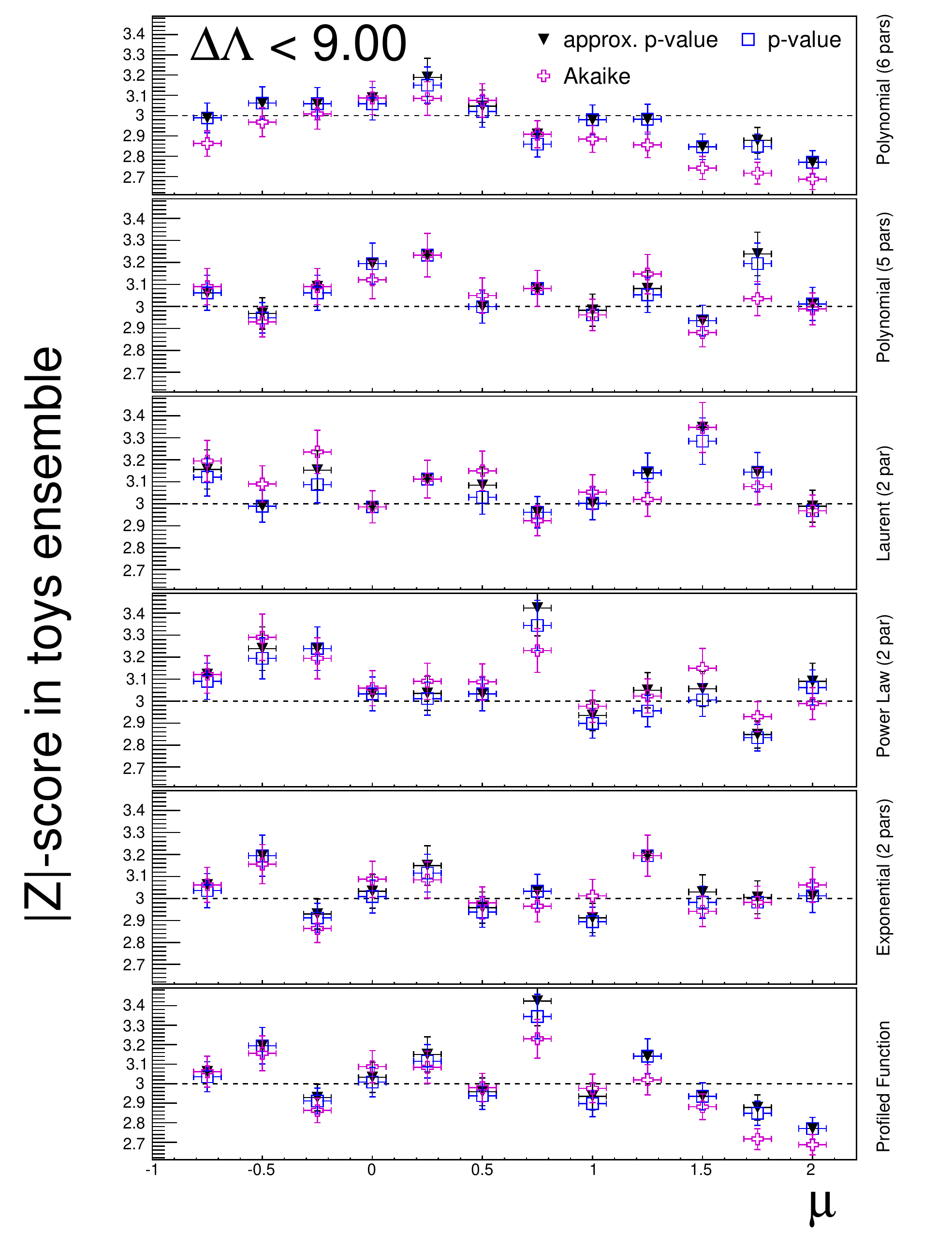}
\label{fig:correction:allorderecoverage:d}
}
\caption{
Measure of the fraction of the toys which do not contain the generated
$\mu$ in various specified $\Lambda$ intervals, converted
into the two-sided score $|Z|$,
when correcting
with the different \nll correction schemes.
The intervals are
$\Delta\Lambda$ of 0.25 (a), 1.0 (b), 4.0 (c) and 9.0 (d).
The results when generating with different functions and generating using the
profiled function for each value of $\mu$ are shown in the different panels.
The functions
used are (from the top panel in each case): polynomial with $N_{\rm par}=6$,
polynomial with $N_{\rm par}=5$, Laurent with $N_{\rm par}=2$,
power law with $N_{\rm par}=2$, and exponential with $N_{\rm par}=2$.
The lowest panel in each case shows the result when the best-fit function at each
value of $\mu$ is used to generate toys.
}
\label{fig:correction:allordercoverage}
\end{figure}

Similarly, figure~\ref{fig:correction:allordercoverage} shows the fraction of
times the envelope fit result was within the various \nll ranges, as used
previously in figure~\ref{fig:functions:firstordercoverage}. Again, the
coverage is reasonably good for all cases. As for the bias, the two
p-value corrections are very similar and always give better coverage than the
Akaike correction.

Hence, overall the two p-value correction methods seem to give effectively
identical results and also perform better than the Akaike correction method.
In practical terms, the approximate p-value correction method is easier
to implement than the exact method
and also can be applied to an unbinned fit (where the $\chi^2$
is not approximately given by the \nll value). For these reasons, the approximate
p-value correction was used throughout the CMS Higgs
to two photon analysis~\cite{ref:introduction:legacy}.

\begin{figure}[tbp]
\centering
\includegraphics[width=0.45\textwidth]{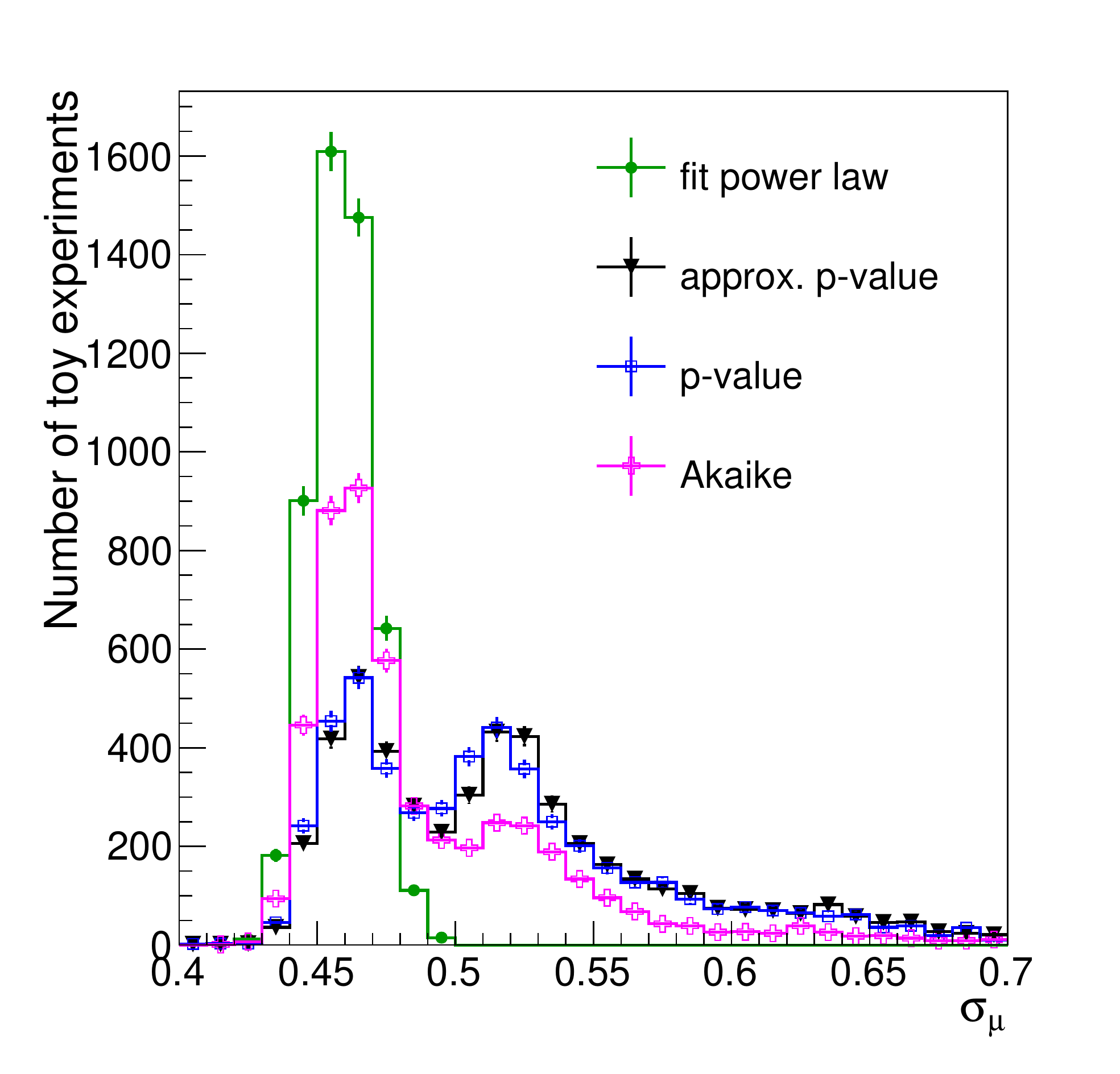}
\caption{Distribution of $\sigma_{\mu}$ in toy datasets
comparing the three correction schemes
when including the full set of functions
with the case when considering only a single power law function
with $N_{\rm par}=2$ for the fit.}
\label{fig:correction:compareerrors}
\end{figure}

Figure~\ref{fig:correction:compareerrors} shows the distribution of the 68.3\%
 uncertainty on $\mu$
when considering the full set of functions using the three different correction schemes compared
with considering a single first order function. The uncertainty ($\sigma_{\mu}$)
is defined as half of the 68.3\% interval on $\mu$.
The distributions are obtained from fitting
toy datasets generated using the first order power law function for the background, with
signal generated assuming $\mu=1$. This quantifies the increase in the error
due to the systematic contribution arising from the uncertainty in the choice of functional form.
The second peak in the distributions observed
when considering the envelope of functions are due to cases in which functions other than the
best fitting function contribute to the 68.3\% interval. As expected, correcting using a higher
penalty reduces the size of this tail since in this case, it is less likely that a higher order
function can contribute.

\subsection{Comments on the choice of correction}
\label{sec:correction:comments}
 As mentioned at the end of Section~\ref{sec:correction:example},
the correction can be considered to have the form
\begin{displaymath}
\Lambda_{\mathrm{corr}} = \textrm{\nll} + cN_{\rm par},
\end{displaymath}
giving a continuum of possible values of $c$, not just $c=1$ or 2.
Figure~\ref{fig:correction:allordercoverage} indicates that in most cases, the
coverage is effectively independent of the value of $c$. This means that,
within a reasonable range,
the value of $c$ used can be motivated by other considerations. In general,
the choice of a particular value of $c$ to use depends on the application
and the amount of data available.

\begin{figure}[tbp]
\centering
\subfigure[]{
 \includegraphics[width=0.46\textwidth]{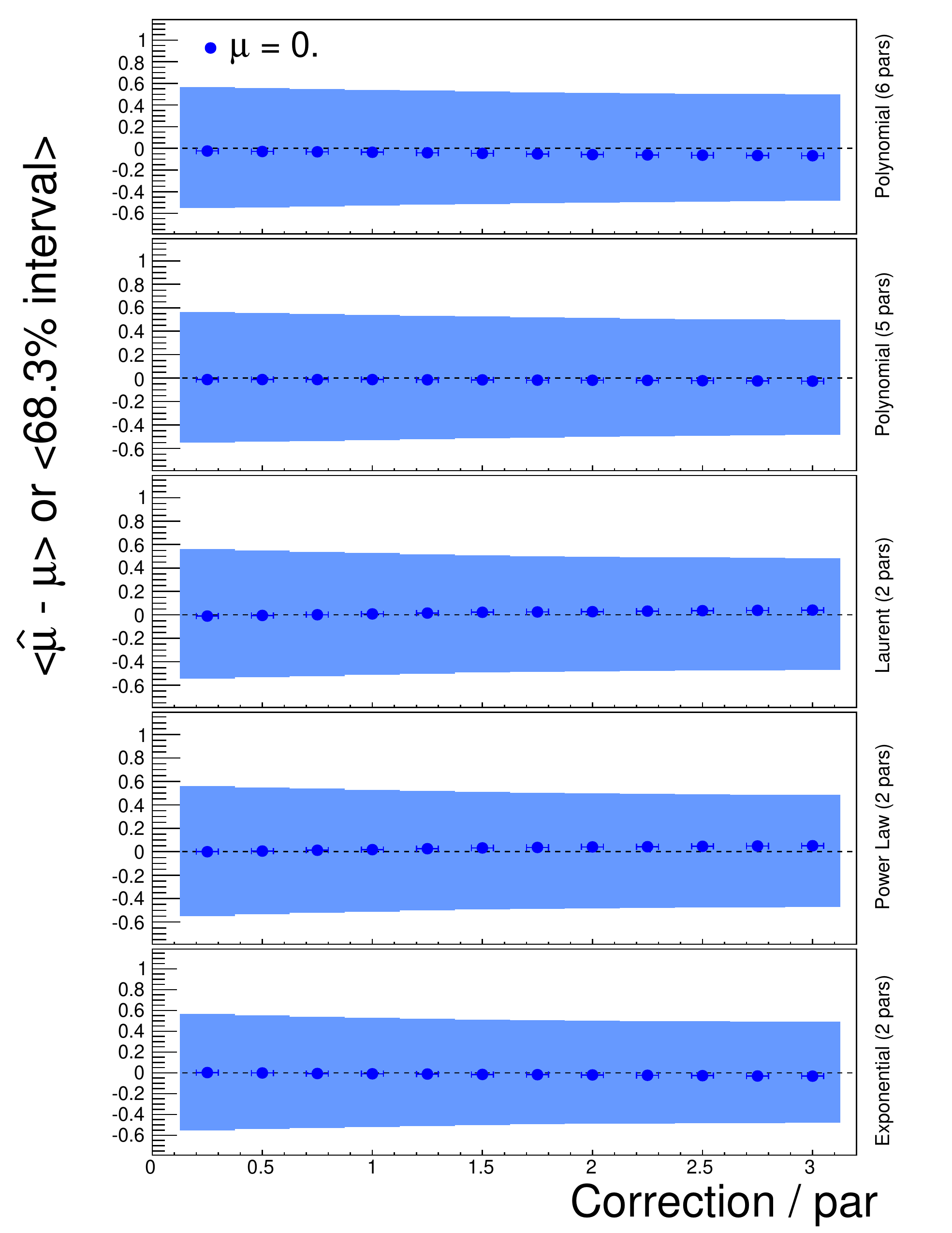}
\label{fig:correction:allordererrvscorr-0}
}
\subfigure[]{
\includegraphics[width=0.46\textwidth]{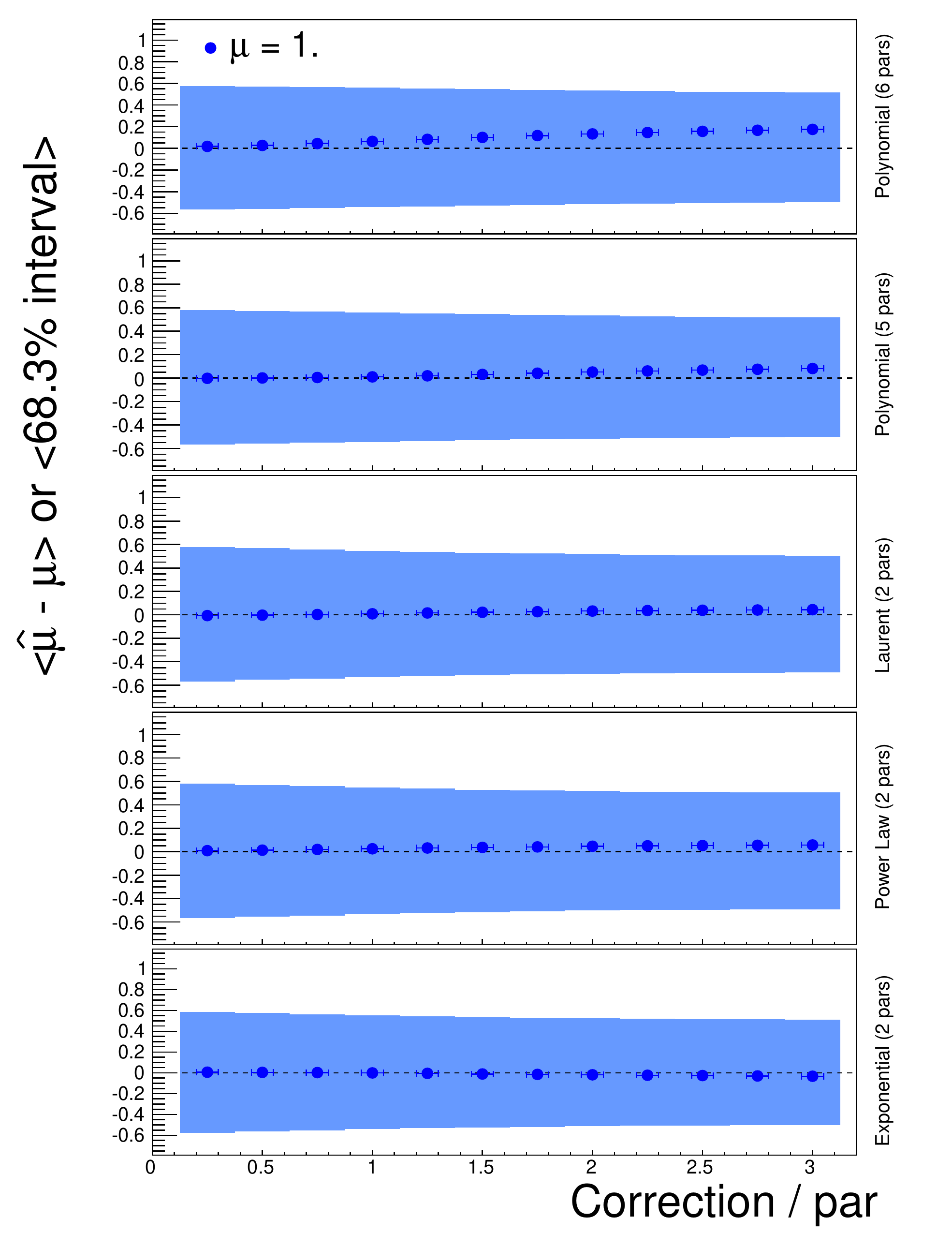}
\label{fig:correction:allordererrvscorr-1}
}
\caption{Mean value of the difference between the fitted and true values
of $\mu$  (data points)
and 68.3\% uncertainty (bands)
as a function of the correction $c$ applied to \nll, for true $\mu=0$ (a)
and $\mu=1$ (b), calculated in toy datasets.
The results when generating with different functions
are shown in the different panels.
The functions
used are (from the top): polynomial with $N_{\rm par}=6$,
polynomial with $N_{\rm par}=5$, Laurent with $N_{\rm par}=2$,
power law with $N_{\rm par}=2$, and exponential with $N_{\rm par}=2$.
}
\label{fig:correction:allordererrvscorr}
\end{figure}

Figure~\ref{fig:correction:allordererrvscorr} shows the mean of the difference
between the best fit $\mu$ value and the true $\mu$, and also the 68.3\%
statistical uncertainty on $\mu$ from the fit, calculated from the toy
samples discussed above, as a function of $c$,
for generated values of $\mu=0$ and 1. It is seen that the $\mu$ offset
is generally small but increases with the correction value $c$. This is
because lower order functions are favoured for large $c$ and these are not as good at
describing more complicated shapes. Note the largest
offset is found when generating with the six-parameter polynomial, as there
are few other functions being used in this study which have a similar number
of parameters.
Conversely, the
statistical error gets smaller for large $c$ as there tend to be fewer parameters
in the functions that contribute to the envelope.

Clearly there is a trade off between statistical power and systematic bias. Furthermore, the
size of this effect is somewhat hidden by the fact that the highest order functions considered
in this example only go up to six free parameters. If even higher order functions were considered
the statistical uncertainty at low values of $c$ would become unnecessarily large.

In principle, the results of figure~\ref{fig:correction:allordererrvscorr}
could be used to choose an ``optimal'' value of $c$
for a given dataset. Here, optimal means in some way trying to minimise
the overall uncertainty, which arises from some combination of
the bias and the statistical error.
It is clear that such an optimal value will change with the
amount of data as the
statistical errors will become smaller but the offset will stay the same and
hence lower values of $c$ will be favoured for larger data volumes.
However, defining how to optimise the value is somewhat arbitrary. It depends
on whether accurately knowing the true uncertainty on $\mu$ is critically
important. The statistical error will be well known but the bias depends on the
underlying background shape which is by definition not well known. Hence,
how to weigh the relative importance of these two contributions is not uniquely
defined. For example, the offset and statistical error could be added in
quadrature to give an estimate of the overall error. For the case of
figure~\ref{fig:correction:allordererrvscorr}, this has its smallest
value at high $c$. However, for the actual Higgs to two photon analysis, the value of $c=1$
was chosen. This is more conservative, in that it means the known statistical
error will dominate.

A decision on the value of $c$ will therefore depend on the application and
no absolute method for determining this can be given. In general a very small
value of $c$ will result in a larger statistical uncertainty and a large value will result in a larger systematic bias.

%% file: conclusions.tex
\section{Discussion and conclusions} 
\label{sec:conclusions}
The actual CMS
Higgs to two photon analysis~\cite{ref:introduction:legacy}
is significantly more
complex than the simpler method discussed above. In particular, the 2011
and 2012 data
samples are split into eleven and fourteen categories, respectively, which have
differing signal to background ratios.
Because the categories (by definition) have different selection criteria,
they can have different background shapes.
There is no {\it a priori} reason to make any assumptions that the functions
used in each category should be the same. Hence, each category should be
tested with all functions, in a similar way to the above.

A major complication then arises because there are common systematic effects
across the categories, arising from nuisance parameters in the signal
model.
In the absence of these common nuisance parameters,
the different categories could be profiled independently, each using the
minimum envelope technique to produce an envelope curve per category.
These could then
be summed to give the overall profile curve. However, with common
nuisance parameters, all categories must be profiled at the same time.
Since minimisation code to handle the
discrete nuisance parameter identifying the
function seems difficult, in practical terms, this means that all possible
combinations of each function in each category must be fitted.
The minimum envelope made from the results of all these combinations would
then be found. While this is conceptually straightforward, the actual
naive implementation is prohibitive and approximations were taken in practice,
while retaining the core of the method.

In conclusion,
a method of treating the uncertainty due to the choice of background function
as a discrete nuisance parameter has been shown to give good coverage and
small 
bias.
Although described in terms of a particular application, the method can be
applied very widely as the general type of problem for which this method
is relevant is very common.

There is an uncertainty about how to penalise functions with higher numbers
of parameters. The results indicate that a penalty based on the p-value
is less biased and gives slightly better coverage than one based on the
Akaike criterion. In general, the choice of the size of the correction must be determined
on a case-by-case basis. The approximate p-value correction is easier to implement and
can be applied more widely and so was chosen as the method to use for the CMS Higgs to
two photon analysis.
In the studies presented, the actual results are relatively independent of the exact value of the penalty
applied.

